\documentclass[iop,twocolappendix]{emulateapj}
\pdfoutput=1
\usepackage{textcase}
\usepackage{amssymb,amsmath}
\usepackage{graphicx}
\usepackage{float}
\usepackage{multirow}
\usepackage{enumerate,mdwlist}
\usepackage{rotating}
\usepackage{tablefootnote}
\usepackage{color}
\usepackage[backref,breaklinks,colorlinks,citecolor=blue]{hyperref}
\usepackage[all]{hypcap}
\newcommand{\jpl}{2}
\newcommand{\caltech}{1}
\newcommand{\boulder}{3}
\newcommand{\penn}{4}

\shorttitle{The Cloverleaf with \emph{Herschel} Spectroscopy}
\shortauthors{Uzgil et al.}

\begin{document}

\title{Constraining the ISM properties of the Cloverleaf quasar host galaxy with \emph{H\MakeTextLowercase{erschel}} spectroscopy}

\author{Bade~D.~Uzgil\altaffilmark{\caltech}}

\author{C.~Matt~Bradford\altaffilmark{\caltech,\jpl}}

\author{Steve~Hailey-Dunsheath\altaffilmark{\caltech}}

\author{Philip R. Maloney\altaffilmark{\boulder}}

\author{James~E.~Aguirre\altaffilmark{\penn}}

\email{badeu@caltech.edu}

\altaffiltext{\caltech}{Division of Physics, Math and Astronomy, California Institute of Technology, Pasadena, CA 91125}

\altaffiltext{\jpl}{Jet Propulsion Laboratory, Pasadena, CA 91109}

\altaffiltext{\boulder}{Center for Astrophysics and Space Astronomy, University of Colorado, Boulder, CO 80309}

\altaffiltext{\penn}{Department of Physics and Astronomy, University of Pennsylvania, Philadelphia, PA 19104}

\begin{abstract}

We present \emph{Herschel} observations of far-infrared (FIR) fine-structure (FS) lines [CII]158$\mu$m, [OI]63$\mu$m, [OIII]52$\mu$m, and [SiII]35$\mu$m in the $z=2.56$ Cloverleaf quasar, and combine them with published data in an analysis of the dense interstellar medium (ISM) in this system. Observed [CII]158$\mu$m, [OI]63$\mu$m, and FIR continuum flux ratios are reproduced with photodissociation region (PDR) models characterized by moderate far-ultraviolet (FUV) radiation fields $G_0=$ 0.3--1$\times10^3$ and atomic gas densities $n_{\rm H}=$ 3--5$\times10^3$~cm$^{-3}$, depending on contributions to [CII]158$\mu$m from ionized gas. We assess the contribution to  [CII]158$\mu$m flux from an active galactic nucleus (AGN) narrow line region (NLR) using ground-based measurements of the [NII]122$\mu$m transition, finding that the NLR can contribute at most 20--30\% of the observed [CII]158$\mu$m flux. The PDR density and far-UV radiation fields inferred from the atomic lines are not consistent with the CO emission, indicating that the molecular gas excitation is not solely provided via UV-heating from local star-formation, but requires an additional heating source. X-ray heating from the AGN is explored, and we find that X-ray dominated region (XDR) models, in combination with PDR models, can match the CO cooling without overproducing observed FS line emission. While this XDR/PDR solution is favored given the evidence for both X-rays and star-formation in the Cloverleaf, we also investigate alternatives for the warm molecular gas, finding that either mechanical heating via low-velocity shocks or an enhanced cosmic-ray ionization rate may also contribute. Finally, we include upper limits on two other measurements attempted in the \emph{Herschel} program: [CII]158$\mu$m in FSC~10214 and [OI]63$\mu$m in APM~08279+5255.
\end{abstract}

\keywords{far-infrared spectroscopy; individual galaxies (H11413+117); interstellar medium; feedback; active galaxies}

\section{Introduction}

Parallel histories of cosmic star formation (SF) and supermassive black hole (SMBH) accretion are suggestive of a causal relationship between the two processes, yet the nature of this link remains an open question in astrophysics.  At the root of this connection is the cold molecular gas in galaxies, which must be shared as fuel for both growing black holes and building stellar nurseries.  Far from simple competitors, however, the roles of SF and SMBH growth in a galaxy's evolution are varied and complex. (See, e.g., reviews on the subject by  \citet{Heckman2014} and \citet{Madau2014}). Molecular, star-forming gas in the circumnuclear region of galaxies known to host accreting SMBHs (called Active Galactic Nuclei, or AGN) are particularly useful test-beds for theories relating the feedback of the SMBH on SF (and vice versa) given the relatively short distances ($\sim1$ kpc) between the molecular gas and the SMBH. 

The $z=2.56$ Cloverleaf quasar and its host galaxy have emerged as a case study for co-occurring SF and SMBH accretion during the epoch of peak galaxy assembly. Although the Cloverleaf was initially discovered in an optical survey of luminous quasars \citep{Hazard1984}, follow-up observations \citep{Barvainis1992} of the sub-millimeter (submm) continuum revealed an excess in the rest-frame far-infrared (FIR) portion of its spectral energy distribution (SED) that was consistent in spectral shape with thermal emission from dust. This fact and the knowledge that the quasar is strongly gravitationally lensed \citep{Magain1988}, combined with the advent of high-$z$ CO measurements \citep{Brown1991}, rendered the Cloverleaf a prime target for CO line searches. The detection of CO($J=3\rightarrow2$) in the system \citep{Barvainis1994} effectively launched the Cloverleaf into the early stages of its longstanding role as a laboratory for high-$z$ studies of molecular gas and SF in the environs of a powerful AGN. Since the first successful CO measurement, the Cloverleaf has been observed, to date, in numerous tracers of molecular gas, including 8 transitions of the CO ladder ($J=1\rightarrow0$ \citep{Riechers2011clover}, $3\rightarrow2$ \citep{Barvainis1997,Weiss2003}, $4\rightarrow3$, $5\rightarrow4$ \citep{Barvainis1997}, $6\rightarrow5$ \citep[hereafter B09]{Bradford2009}, $7\rightarrow6$ \citep[B09]{Alloin1997,Barvainis1997}, $8\rightarrow7$, and $9\rightarrow8$ (B09)), two fine-structure (FS) transitions of [CI] ($^3$P$_{1} \rightarrow ^3$P$_{0}$ \citep{Weiss2005} and $^3$P$_{2} \rightarrow ^3$P$_{1}$ \citep{Weiss2003}), HCN ($J=1-0$)\citep{Solomon2003}, and HCO$^+$ ($J=1\rightarrow0$ \citep{Riechers2006hco} and $4-3$ \citep{Riechers2011hco43}), and CN \citep{Riechers2007cn}. Spatial extent of the molecular gas, derived from a CO($J=7\rightarrow6$) map, has also been assessed, and appears to be concentrated in a disk of radius 650 pc, centered on the SMBH \citep[VS03]{VS2003}. Non-LTE modeling of CO excitation with an escape probability formalism suggests that all observed transitions can be described by a single gas component \citep{Bradford2009, Riechers2011clover}, so there is no indication of significant molecular emission in the observed lines beyond the CO($J=7\rightarrow6$) disk. Physical conditions inferred from the modeling point to $n_{\rm H_2}=2$--3$\times10^4$ cm$^{-3}$ and $T=50$--60~K, suggesting that the CO gas is distributed uniformly or with high areal filling factors---not in sparse clumps---in order to maintain this thermal state throughout the $\sim1$~kpc-wide emitting region. 

In addition to molecular spectroscopy, (sub)mm continuum measurements have provided further insight into the nature of SF in the Cloverleaf ISM. In the rest-frame IR SED compiled by \citet{Weiss2003}, the Cloverleaf's continuum emission appears double-peaked, with distinct cold and warm gas components with dust temperatures of $\sim50$~K and $\sim115$~K, respectively. The starburst origin of the cold gas component is strongly supported by the detection of emission features from polycyclic aromatic hydrocarbons (PAHs) in the Cloverleaf's rest-frame mid-infrared spectrum, which were shown to follow the empirical correlation to FIR luminosity established for starbursts and composite quasar/starburst systems in the local Universe \citep{Lutz2007}. Attributing, then, the entirety of the FIR (42.5--122.5\rm$\mu$m) luminosity, $L_{\rm FIR}$, inferred from the cold component of the SED, reveals a starburst of intrinsic $L_{\rm FIR} = 5.4\times10^{12}$~L$_{\odot}$. 

The presence of a major starburst suggests that molecular gas in the Cloverleaf is heated, at least in part, within the depths of photodissociation regions (PDRs) illuminated by far-ultraviolet photons from the ongoing star formation. However, while categorically ``ultraluminous," the Cloverleaf starburst accounts for only 5--10\% of the total bolometric luminosity, $\sim7\times10^{13}$~L$_{\odot}$; the system energetics remain dominated by the central engine. In fact, B09 estimate an X-ray flux toward the edge of the 650 pc radius of $\sim10$~erg~s$^{-1}$~cm$^{-2}$, and find that irradiation by X-rays could play a significant role in gas heating, as well. They explain the high total CO-to-far-IR continuum ratio of $6\times10^{-4}$, atypical of local starbursts, with a scenario in which X-ray dominated regions (XDRs) contribute substantially to the CO emission.

Identifying the dominant heating source of the molecular gas in the Cloverleaf is essential to understanding the relationship between the SMBH and SF in the host galaxy and, particularly, the role of AGN in regulating star-formation. In this paper, we present new measurements of key diagnostic lines of atomic and ionized media to aid in the interpretation of the excitation mechanisms for the observed CO in the Cloverleaf disk. The detected lines, namely [CII]158$\mu$m, [OI]63$\mu$m, [OIII]52$\mu$m, and [SiII]35$\mu$m, provide highly complementary information to the CO spectroscopy by tracing star-forming gas in different phases of the ISM, and by providing additional means to test XDR and PDR models, which can predict bright emission in the observed atomic lines.

This article is organized as follows. First, we report in Section~\ref{sec:observations} the measured line fluxes from \emph{Herschel}-SPIRE and -PACS instruments, and discuss uncertainties where necessary. With observations of the important PDR cooling lines [CII]158$\mu$m and [OI]63$\mu$m enabled by \emph{Herschel}, we are able to infer the average densities and FUV fluxes prevalent in the Cloverleaf PDRs by employing traditional FS line ratio diagnostics, as well as to better estimate the relative contribution of the AGN and SF to producing the observed emission, which we explore in Section~\ref{sec:analysis}. There, after subtracting contributions from ionized gas in the Narrow Line Region (NLR) and HII regions, we compare measured line ratios of the FS lines and CO to predicted values from PDR and XDR models and determine their respective contributions to the observed emission. We also briefly consider shock excitation of CO as an alternative explanation for the unusual high total CO line-to-FIR continuum ratio. Finally, in Section~\ref{sec:Discussion}, we place our findings for the Cloverleaf in the context of other AGN discovered at similar and lower redshifts.

\section{Observations} \label{sec:observations}

Measured fluxes for the fine-structure (FS) lines obtained in this work are presented in Table~\ref{tab:flux_table}. We supplement our measurements with published fluxes for the [NII]122$\mu$m emission, CO up to $J_{upper}=1$--9, and the 6.2~$\mu$m and 7.7~$\mu$m PAH emission features.

\paragraph{SPIRE FTS}

\begin{figure*}
\centering
\begin{tabular}{c}
\includegraphics[width=0.8\textwidth]{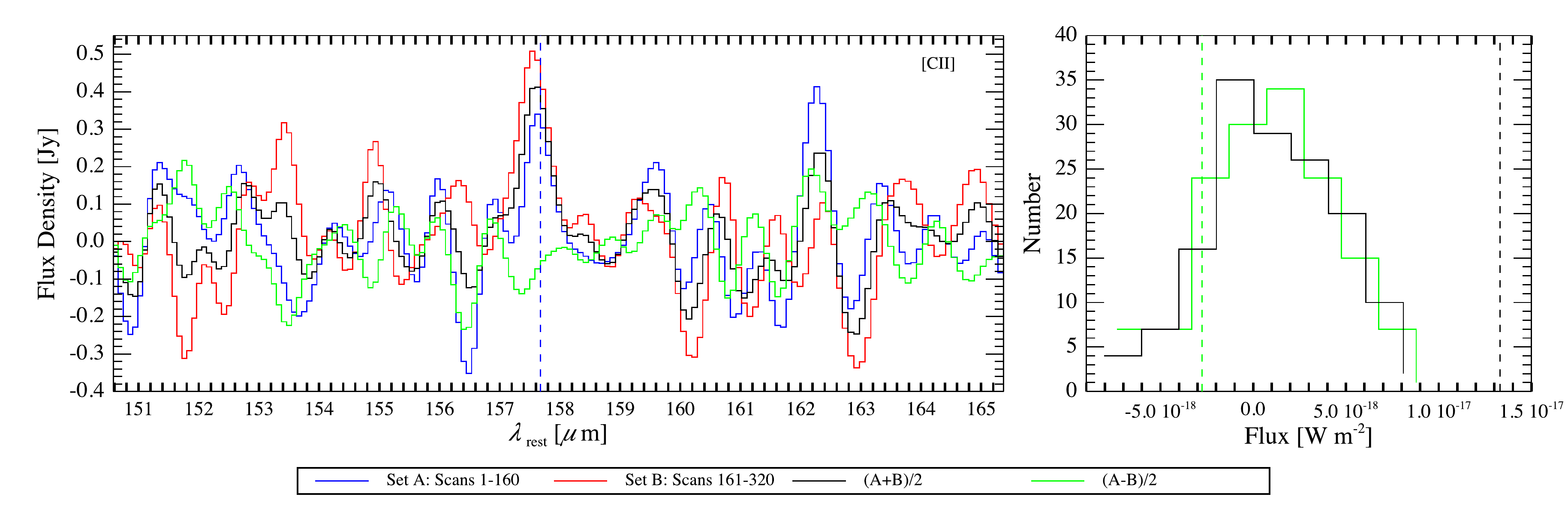} \\
\includegraphics[width=0.8\textwidth]{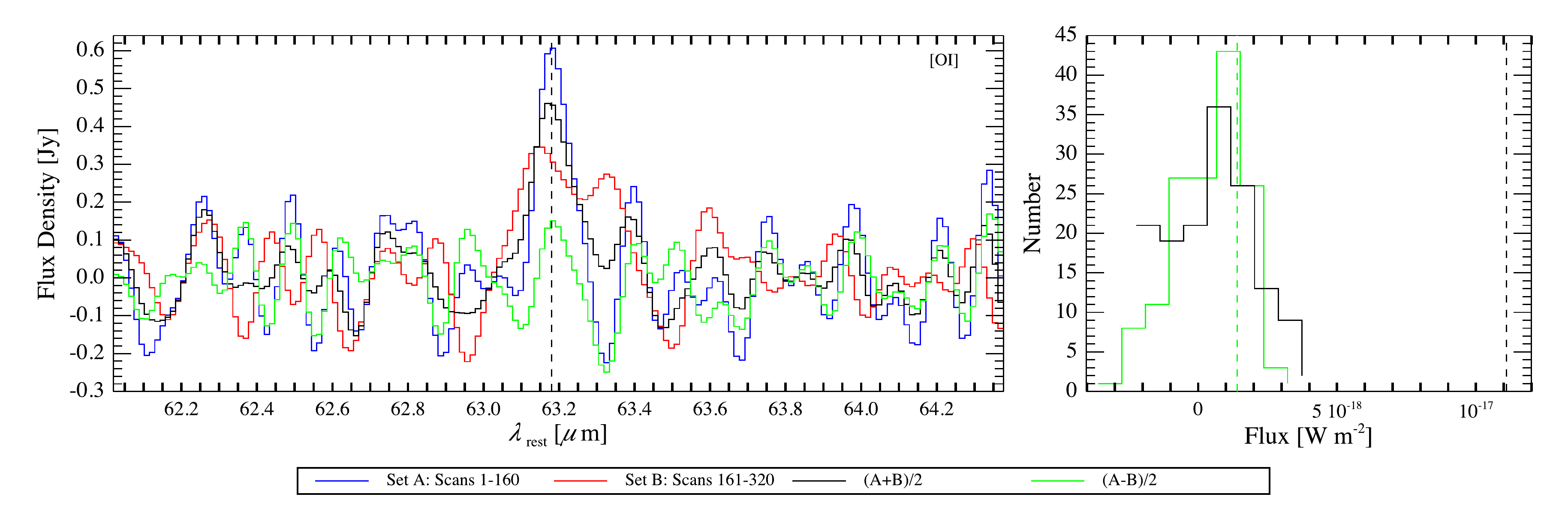} \\
\includegraphics[width=0.8\textwidth]{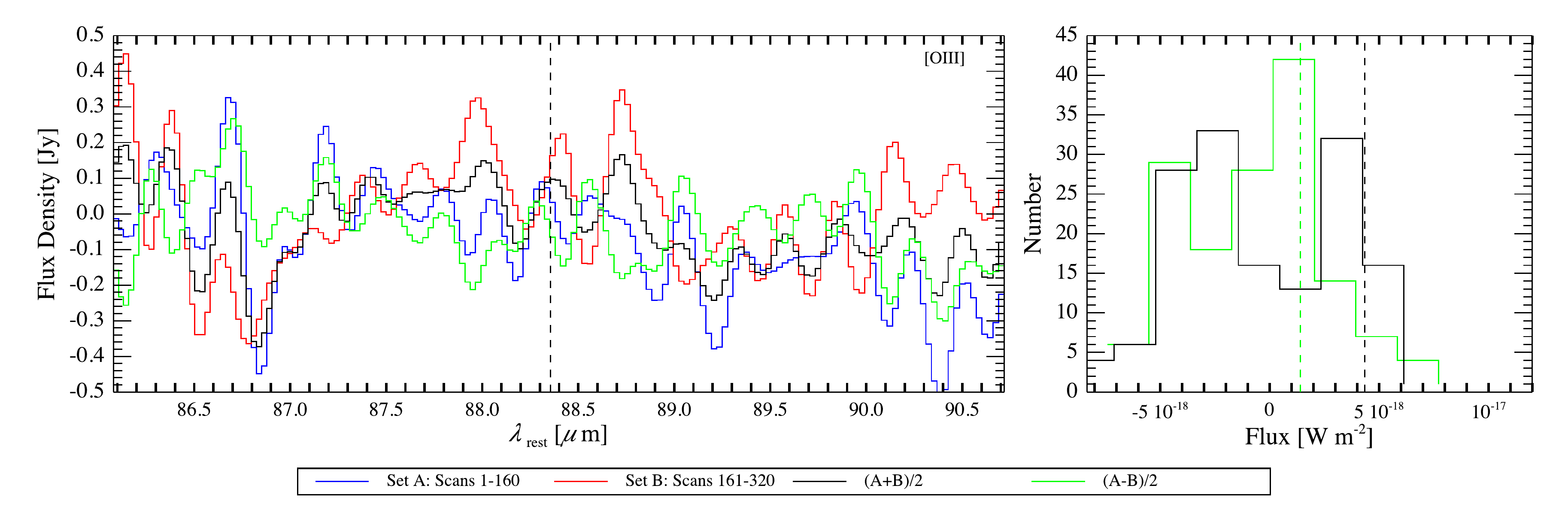} \\
\end{tabular}
\caption{SPIRE FTS spectra for [CII]158$\mu$m, [OI]63$\mu$m, and [OIII]88$\mu$m (top to bottom panels). Each spectrum shown here spans a 50~GHz range (in the observed frame) centered at the rest wavelength for the target line. Blue and red curves denote spectra from the first half of scans (i.e., scan numbers 1 through 160) and second half of scans (scan numbers 161 through 320) in the dataset, while black and green curves correspond to the coadded and jackknifed data. Histograms at right show the flux returned by the line fit applied to each frequency position in the 50~GHz ranges of the coadded (black) and jackknifed (green) spectra, while avoiding the surrounding $2\times$FWHM line widths on each side of the target line center. Black and green dashed vertical lines denote the fitted flux at the target line center for the coadded spectrum and the jackknifed spectrum, respectively.}
\label{fig:jackknives_hist}
\end{figure*}

At $z = 2.56$---the CO redshift of this source---fine-structure line emission from [CII]158$\mu$m and [OI]63$\mu$m is redshifted to within coverage of the long wavelength (LW, 303--671~$\mu$m) and short wavelength (SW, 194--313~$\mu$m) bands of the SPIRE Fourier Transform Spectrometer (SPIRE FTS) aboard \emph{Herschel}. Point source spectra were obtained in sparse observing mode for the Cloverleaf with a total of 320 FTS scans---160 in each forward and reverse directions---from the \emph{Herschel} OT program \texttt{OT1\_mbradfor\_1} (PI: Matt Bradford). Amounting to 364.4 minutes of observing time for the source, these spectra are the deepest SPIRE spectra yet presented, to our knowledge. The continuum level is at 0.1--0.5~Jy, which is close to the continuum flux accuracy achieved on SPIRE. As such, we take care to address concerns about spurious line detections arising from random noise fluctuations in the continuum, and---once lines have been identified---to accurately quantify uncertainties in the measured line fluxes.

To reduce the probability of a spurious line detection, we perform a jackknife test for each targeted line in which the full set of 320 unapodized spectra obtained from corresponding FTS scans is first split into two subsets. The jackknife split we apply is temporal, in order to test for variations in the spectrum as a function of observing time; we simply divide the scan set into halves containing scans 1--160 and 161--320, where scan 1 denotes the beginning of the observation and 320, the end. The 160 spectra in each half are then co-added to produce two separate spectra (called A and B), and then differenced to produce a residual spectrum. In the absence of systematic error between sets A and B, the differenced spectrum will contain zero flux at all wavelengths. Figure~\ref{fig:jackknives_hist} shows the results of the jackknife tests for the [CII]158$\mu$m, [OI]$\mu$m63, and [OIII]88$\mu$m lines in 50~GHz segments centered at the rest wavelength for each line. Note that the 50~GHz bandwidths are computed in the observed frame.

Line fluxes were measured from the unapodized spectrum using the SpectrumFitter in \emph{Herschel} Interactive Processing Environment (HIPE) 12 version 1.0 \citep{Ott2010}. We fit each emission line and 6~GHz (observed frame) of local continuum\footnote{For reference, 6~GHz of local continuum corresponds to velocity widths of 3,360~km~s$^{-1}$, 1,350~km~s$^{-1}$, and 1,890~km~s$^{-1}$ in spectra containing [CII]158$\mu$m, [OI]63$\mu$m, and [OIII]88$\mu$m, respectively.} with a 1st order polynomial baseline and a Gaussian line profile convolved with a sinc function (i.e., ``SincGauss" model in HIPE). Line centers, $(1+z_{\rm Clover})\lambda_{\rm rest}$ were fixed, as were the widths of the Gaussian and sinc profiles. The assumed Gaussian widths are not crucial to the fit---we adopted a FWHM of 500~km/s, on the upper range of that measured by \citet{Weiss2003} in the CO lines with the Plateau de Bure interferometer. For the sinc function, we fixed the width at 0.38~GHz, which is set by the 1.2~GHz spectral resolution\footnote{This frequency resolution corresponds to velocity resolutions 670~km~s$^{-1}$, 270~km~s$^{-1}$, and 380~km~s$^{-1}$, respectively, for spectra containing [CII]158$\mu$m, [OI]63$\mu$m, and [OIII]88$\mu$m.} of the FTS (in high resolution mode). The SpectrumFitter returns the fitted parameters for the SincGauss model, which we convert to flux using the appropriate analytic formula. It also provides an associated uncertainty, which we consider as a lower limit. We proceed to generate our own estimate of the RMS noise in 50~GHz of bandwidth centered at the target frequency. In our estimate, we repeatedly fit our SincGauss model at various frequencies within this bandwidth. Because the SPIRE FTS has a spectral resolution of 1.2~GHz, there are 41 independently sampled frequencies---and thus 41 independent line flux fits---in each 50~GHz range. The uncertainty reported in Table~\ref{tab:flux_table} is the standard deviation of the sampled frequencies, plus a 5\% calibration uncertainty for SPIRE. A histogram of the flux fits using our uncertainty estimate is displayed in the righthand panels of Figure~\ref{fig:jackknives_hist}. For reference, the measured flux of the target line is also shown (dashed vertical line) in the histogram plot, and is given in Table~\ref{tab:flux_table}. The lines [CII]158$\mu$m and [OI]63$\mu$m are detected at levels of $3.8\sigma$ and $8.5\sigma$, respectively. A $3\sigma$ upper limit is reported for [OIII]88$\mu$m. 

\paragraph{PACS}

\begin{figure*}[t]
\centering
\begin{tabular}{c c}
\includegraphics[width=0.4\textwidth]{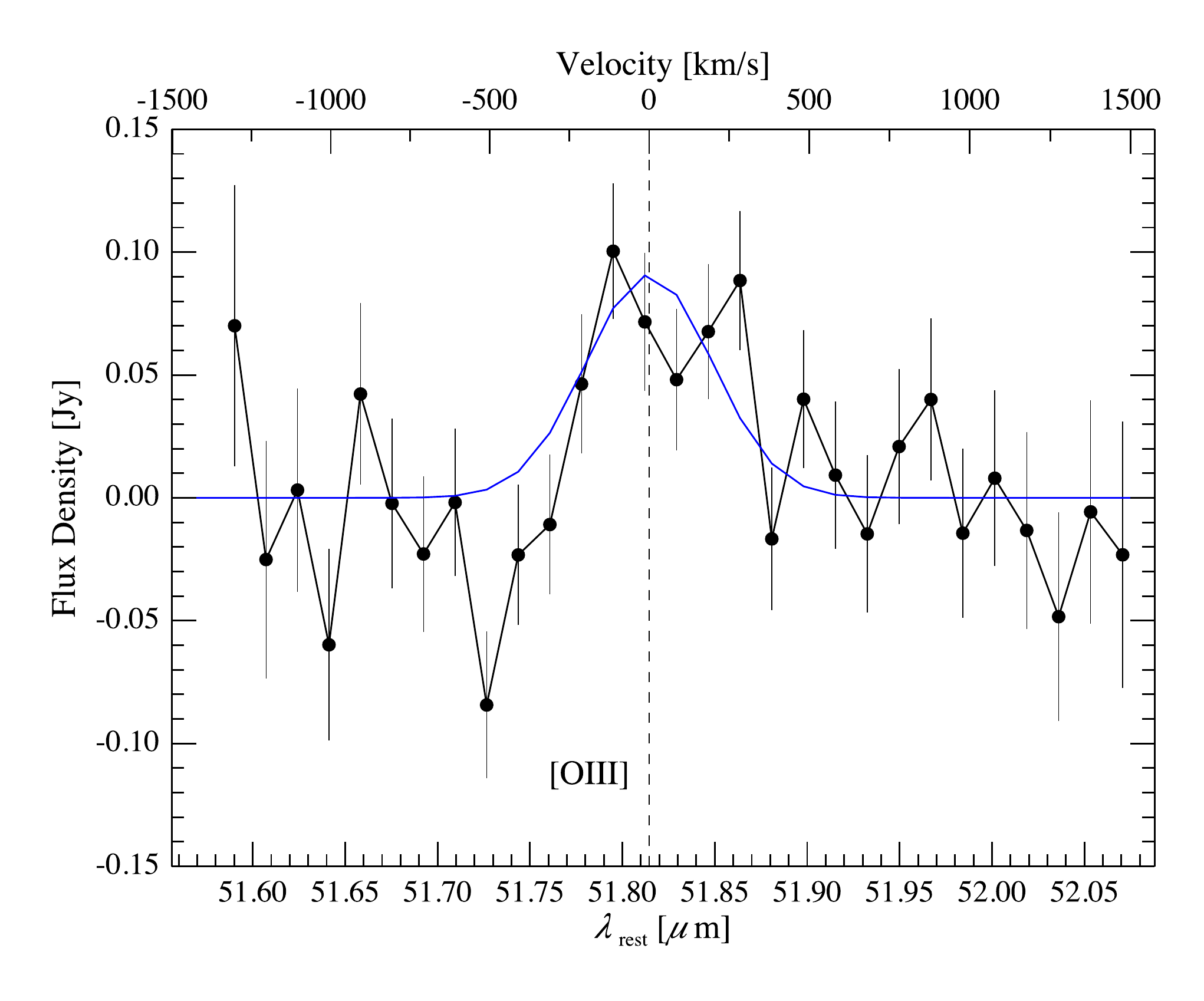} &
\includegraphics[width=0.4\textwidth]{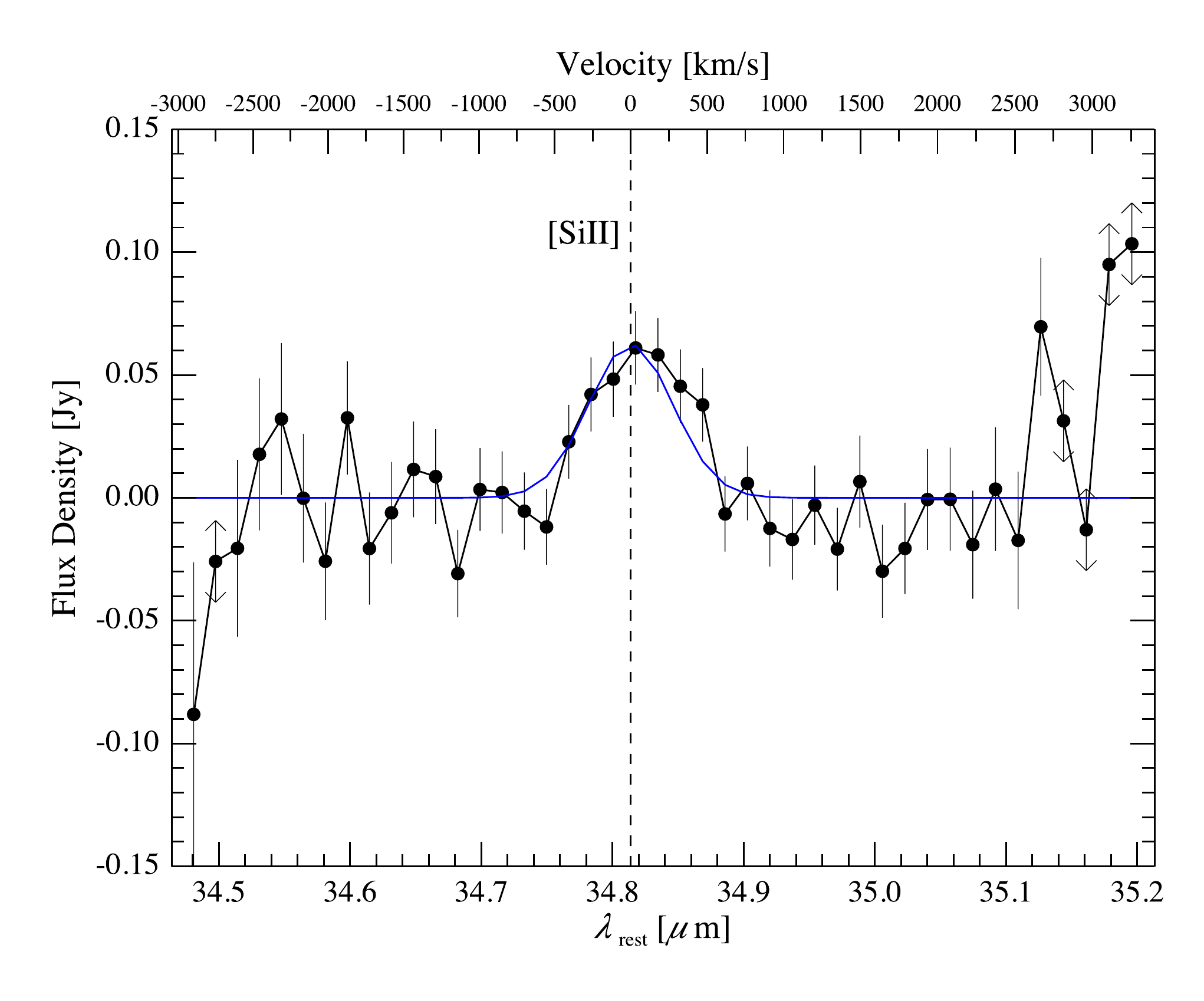} \\
\includegraphics[width=0.4\textwidth]{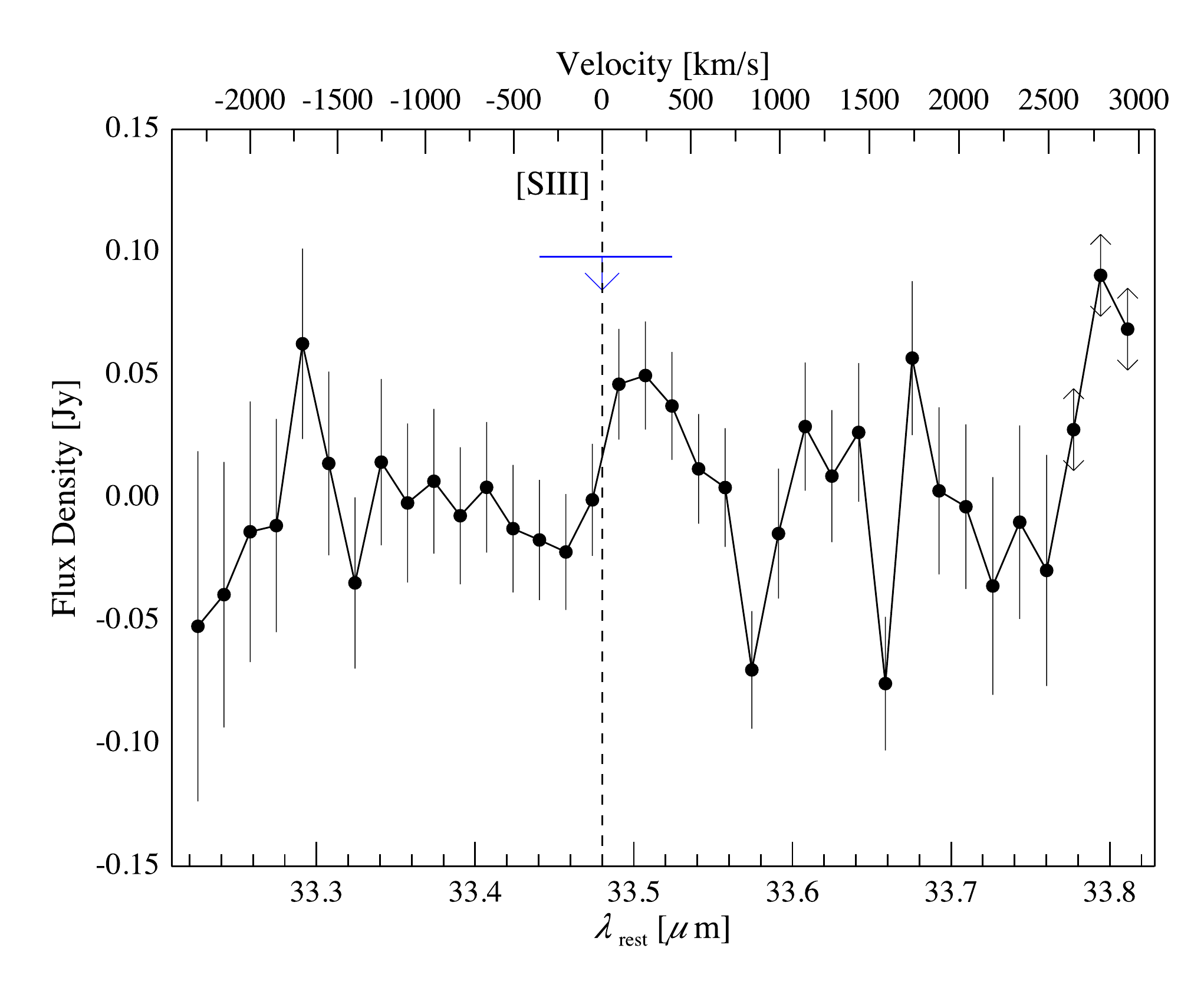} &
\includegraphics[width=0.4\textwidth]{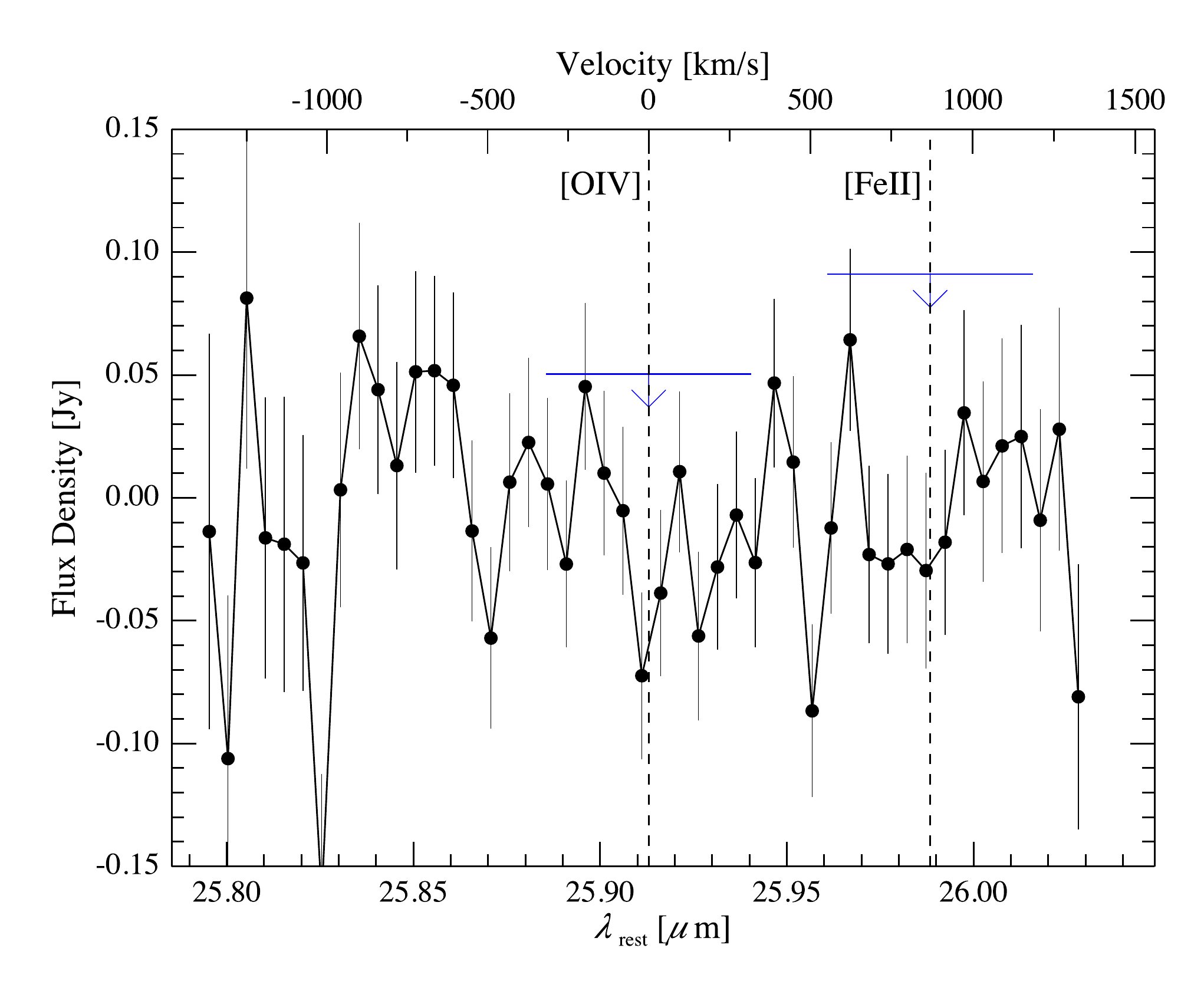} \\
\end{tabular}
\caption{Resampled continuum-subtracted PACS spectra with error bars for [OIII]52$\mu$m, [SIII]33$\mu$m, [SiII]35$\mu$m, and [OIV]26$\mu$m and [FeII]26$\mu$m. Gaussian fits (FWHM = 500 km/s, fixed width) for spectral lines are shown as blue curves. Horizontal blue lines with downward-pointing arrows indicate a $3\sigma$ upper limit on the line flux density.}
\label{fig:pacs}
\end{figure*}

Point-source observations of the Cloverleaf include spectra from the blue and red channels, which cover the wavelength range corresponding to expected FS line emission from [OIII]52$\mu$m, [SiII]35$\mu$m, [SIII]33$\mu$m, [FeII]26$\mu$m, and [OIV]26$\mu$m. Spectra for [OIII]52$\mu$m, [SiII]35$\mu$m, [FeII]26$\mu$m, and [OIV]26$\mu$m were obtained from the \texttt{OT1\_mbradfor\_1} observing program; additional spectra containing [OIII]52$\mu$m, [FeII]26$\mu$m, and [OIV]26$\mu$m, along with a spectrum for [SIII]33$\mu$m, were obtained from the Key Program \texttt{KPGT\_kmeisenh\_1} (PI: Klaus Meisenheimer). Spectra from multiple observations of the same target lines were co-added using the AverageSpectra task in HIPE. The data were reduced with the background normalization version of the chopped line scan reduction script included in HIPE. We reduced the data with $oversample =4$, then binned by a factor of 2 in post-processing to achieve Nyquist sampling. Processed spectra with error bars and line fits (blue curves) for the targeted lines are presented in Figure~\ref{fig:pacs}. Line widths were allowed to vary during each fit, though we kept line centers fixed. Measured fluxes and uncertainties (obtained directly from SpectrumFitter) are listed in Table~\ref{tab:flux_table}. We report detections for [OIII]52$\mu$m (5.1$\sigma$) and [SiII]35$\mu$m (7.8$\sigma$), and upper limits for [OIV]26$\mu$m, [FeII]26$\mu$m, and [SIII]33$\mu$m. The estimated line widths (FWHM) for [OIII]52$\mu$m and [SiII]35$\mu$m are $470\pm150$~km~s$^{-1}$ and $650\pm280$~km~s$^{-1}$, respectively.

\subsection{Extinction corrections}\label{subsec:extinction}
While the far-IR lines are relatively extinction-free, corrections are required for the most highly-obscured systems.   In Arp~220, for example, \citet{Rangwala2011} find that the dust is optically thick at 240$\mu$m, corresponding to a column density in hydrogen, $N_{\rm H}$, of $10^{25}\,\rm cm^{-2}$.   This extreme source demands corrections even for the submillimeter mid-J CO transitions.

The Cloverleaf has similar gas and dust masses to Arp~220, but the extinction is reduced because the size scale is larger.  We estimate extinction values using both gas and dust mass, in both cases spread over the 650-pc radius disk (including the 30$^\circ$ inclination), with an area of 1.1$\times 10^{43}\,\rm cm^2$.  For the gas mass, we take the peak of the B09 molecular gas mass likelihood of $6\times10^9\,\rm M_{\odot}$, which corresponds to a typical hydrogen column in the disk of $4.7\times10^{23}\,\rm cm^{-2}$.   Per the mixed-dust model of \citet{LiDraine2001}, this column creates 0.8 magnitudes of extinction at 63$\mu$m, corresponding to an optical depth, $\tau$, of 0.73; the model assumes a $\nu^{-2}$ scaling with $\tau$ for wavelengths between 30$\mu$m and $1000$m. A similar estimate is obtained with the estimated dust mass from \citet{Weiss2003}, some $6.1\times 10^{7} \,\rm M_{\odot}$.  When distributed in the disk, this gives a mass column of $1.1\times 10^{-2}\,\rm g\,cm^{-2}$.  The absorption coefficient in Table~6 of \citet{LiDraine2001}---adjusted to 63 microns---gives $\kappa_{63} = 84.7\rm\, cm^2 g^{-1}$, and an optical depth of 0.93.  We adopt the average of these values ($\tau_{63} = 0.83$), and use a mixed dust extinction model in which the correction factor relating observed to intrinsic flux is $\tau_d / (1 - e^{-\tau_d})$.  The mixed-dust model is appropriate if the emission lines originate in gas mixed approximately uniformly with the disk, as is the basic scenario for the star-forming disk material.    For emission from the AGN narrow-line region (NLR),  the correction could potentially be greater if the NLR gas is fully covered by the disk.  On the other hand, at least some portion of the NLR material is largely unobscured, since NLR gas is visible at optical wavelengths.   Given these uncertainties, the modest correction of the mixed dust model seems appropriate.  The correction factor for [OI]63$\mu$m is 1.47, and the other transitions are corrected similarly assuming $\tau\propto \nu^{-2}$. Line fluxes reddened according the necessary correction factors are listed in Table~\ref{tab:flux_table}.

\begin{table*}[h]
\centering
\caption{Line fluxes and luminosities for the Cloverleaf}
\label{tab:flux_table}
\begin{tabular}{l c c c c c c}
\hline \hline
Line Transition & $\lambda_{\mathrm{rest}}$ & Observed Flux & Error & Corrected Flux\footnote{Line fluxes (excluding upper limits) have been corrected for extinction according to the mixed dust model from \citet{LiDraine2001}, as described in the main text.} & Intrinsic Luminosity\footnote{Luminosities reported here have been downscaled with the appropriate lensing magnification factor, $\mu=11$ \citep{VS2003}.} & Reference\footnote{Abbreviated references include this work (denoted as U16), \citet{Ferkinhoff2011} (F11), \citet{Riechers2011clover} (R11), \citet{Weiss2003} (W03),  \citet{Barvainis1997} (Ba97), \citet{Bradford2009} (B09), and \citet{Lutz2007} (L07).} \\ 
  & [$\mu$m] & [$10^{-18}$ W m$^{-2}$] & [10$^{-18}$ W m$^{-2}$] & [$10^{-18}$ W m$^{-2}$] & [$10^9$ L$_{\odot}$] & \\
\hline
$\mathrm{[CII]}$ $^2$P$_{3/2} \rightarrow ^2$P$_{1/2}$& 157.7 & $13.3$ & $3.5$ & $14.6$ & $18.9$ & U16\\
$\mathrm{[OIII]}$ $^3$P$_{1} \rightarrow ^3$P$_{0}$ & 88.4 & $<12.4$ (99.7\% CL) & - & - & $<16.0$ & U16\\
$\mathrm{[OI]}$ $^3$P$_{1} \rightarrow ^3$P$_{2}$ & 63.2 & $11.1$ & $1.3$ & $16.3$ & $21.2$ & U16\\
$\mathrm{[OIII]}$ $^3$P$_{2} \rightarrow ^3$P$_{1}$ & 51.8 & $2.67$ & $0.52$ & $4.65$ & $6.00$ & U16\\
$\mathrm{[SiII]}$ $^2$P$_{3/2} \rightarrow ^2$P$_{1/2}$ & 34.8 & $3.18$ & $0.42$ & $9.29$ & $12.0$ & U16\\
$\mathrm{[SIII]}$ $^3$P$_{1} \rightarrow ^3$P$_{0}$ & 33.5 & $<4.38$ (99.7\% CL) & - & - & $<5.65$ & U16\\
$\mathrm{[FeII]}$ $^6$D$_{7/2}\rightarrow ^6$D$_{9/2}$ & 26.0 & $<5.24$ (99.7\% CL) & - & - & $<6.77$ & U16\\
$\mathrm{[OIV]}$ $^2$P$_{3/2} \rightarrow ^2$P$_{1/2}$ & 25.9 & $<2.91$ (99.7\% CL) & - & - & $<3.76$ & U16\\
$\mathrm{[NII]}$ $^3$P$_{2} \rightarrow ^3$P$_{1}$ & 121.8 & $2.40$ & $0.40$ & $2.70$ & $3.49$ & F11\\
CO $J=1\rightarrow0$& 2602.17& $0.00150$ & $0.0000432$ & - & $0.00196$ & R11\\
CO $J=3\rightarrow2$ & 866.98 & $0.043$ & $0.0055$ & - & $0.055$ & W03\\
CO $J=4\rightarrow3$ & 650.25 & $0.0912$ & $0.0035$  & - & $0.117$ & Ba97\\
CO $J=5\rightarrow4$ &  520.24 & $0.130$ & $0.0092$ & - & $0.166$ & Ba97\\
CO $J=6\rightarrow5$ & 433.57 & $0.240$ & $0.053$ & - & $0.31$ & B09\\
CO $J=7\rightarrow6$ & 371.66 & $0.343$ & $0.048$& - & $0.44$ & B09\\
CO $J=8\rightarrow7$ & 325.23 & $0.444$ & $0.041$ & - & $0.57$ & B09\\
CO $J=9\rightarrow8$ & 289.13 & $0.406$ & $0.056$ & - & $0.52$ & B09\\
PAH & 6.2 & $1.36$ & - & - & $19.4$ & L07 \\
PAH & 7.7 & $5.54$ & - & - & $78.8$ & L07 \\
\hline
$L_{\mathrm{FIR}}$(40$\mu$m-120$\mu$m) [10$^{13}$ L$_{\odot}$] & &  & & $0.54$ & W03\\
$L_{\mathrm{Bol}}$ [$10^{13}$ L$_{\odot}$] & &  & & $7.0$ & L07\\
$L_{\rm CO}/M_{\rm H_2}$ [L$_{\odot}$ M$_{\odot}$$^{-1}$] & &  & & 0.39 & B09 \\
\hline
\end{tabular}
\end{table*}

\section{Interpretation of the Data} \label{sec:analysis}

In this section, we use the suite of CO rotational transitions (from the literature), the [NII]122$\mu$m line \citep[F11]{Ferkinhoff2011}, and the newly detected ($>3.5\sigma$) atomic FS lines (from this work) to diagnose the physical conditions prevalent in the ISM of the Cloverleaf system. 

Although the dataset of detected lines is rich, the natural complexities of a multiphase ISM must be treated carefully, particularly given that many of the observed transitions can be excited in a variety of physical environments: photon-dominated regions (PDRs), stellar HII regions, diffuse ionized gas of the warm ionized medium, NLR clouds near the AGN, X-ray Dominated Regions (XDRs), or shocks. High resolution spatio-spectral imaging data for the set of observed emission lines is limited, but is helpful in partitioning ISM components when available, as in the case of CO($J=7\rightarrow6$) and [NII]122$\mu$m. The FS line observations presented here are all spatially unresolved; the smallest beam sizes for SPIRE and PACS imaging spectrometers are 17'' and 9'', respectively, compared to the on-sky diameter of 2'' for the optical quasar and the interferometric CO($J=7\rightarrow6$) image. We are therefore considering integrated emission from the composite source and begin by considering how the atomic line emission is partitioned among the various components. 

\subsection{Origins of $\mathrm{[CII]158\mu m}$ emission}

With a first ionization potential (IP = 11.26~eV) lying just below 1~Rydberg, singly-ionized carbon can coexist with both neutral and ionized hydrogen. Thus, while we expect that PDRs illuminated by star-formation may be the dominant source of [CII]158$\mu$m emission, we must consider the ionized gas from SF, and the NLR, as well. (For simplicity, we use the subscript ``HII" to denote ionized gas from SF throughout this analysis.) The sum of each ISM phase's contribution---expressed as the fraction, $\alpha_{\mathrm{[CII]},j}$, of observed flux for [CII]158$\mu$m (or generally any line $i$) and ISM phase $j$---will sum to the measured total line flux, $F_{\rm [CII]}$, so that the flux attributed to PDRs, $F_{\rm [CII],PDR}$, is written as
 \begin{equation}
F_{\rm [CII],PDR} = (1 - \alpha_{\rm [CII],NLR} - \alpha_{\rm [CII],HII}) \times F_{\mathrm{[CII]}} \label{eq:cii_ism}
\end{equation}

\subsubsection{\rm [CII]158$\mu\mathrm{m}$ from non-star-forming gas} \label{subsec:NLR}

As the AGN is responsible for roughly 90\% of the Cloverleaf's bolometric luminosity, we first consider the potential for [CII]158$\mu$m emission arising in the NLR associated with the central engine.

\paragraph{Narrow Line Region} Recent Band 9 observations with ALMA \citep[F15]{Ferkinhoff2015} have imaged the Cloverleaf system in [NII]122$\mu$m, detecting 20\% of the single-aperture flux $F_{\rm [NII]122}=2.4\times10^{-18}$~W~m$^{-2}$ (F11) within a single synthesized beam ($\theta_{\rm ALMA}\sim0.25$") containing the quasar point source. As the authors there explain, the difference in $F_{\rm [NII]122}$ between the ALMA and the single-dish measurements is likely due to the presence of an extended [NII]122$\mu$m-emitting region that has been ``resolved out" by the small ALMA beam; down-weighting or removing data from extended baselines does not improve the signal-to-noise ratio in the ALMA image (cf. Table 2 in F15). Assuming a $5\sigma$ detection requires peak flux densities of 20~mJy per synthesized beam, and that the emission originates from the Cloverleaf disk, we estimate that the unresolved ALMA flux must originate from $r>230$~pc in the source plane, or else it would have been detected. We do not consider the possibility of a large [NII]122$\mu$m-bright NLR component outside the plane of the disk, as this requires spatially resolved data of the total [NII]122$\mu$m flux distribution, which is currently unavailable. This extended component, responsible for $>80$\% of $F_{\rm [NII]122}$, likely originates mainly from star-formation as traced by the spatially resolved distribution of underlying rest-frame 122$\mu$m continuum emission, which spans a $\sim2$" on-sky diameter and encompasses the emission from all four lensed images of the Cloverleaf (F15). Although the observed [NII]122$\mu$m region is not conclusive evidence for AGN-heating and may include emission from a nuclear starburst, we consider the scenario that AGN-heating is the dominant heating mechanism at $r\le90$~pc---the physical extent of $\theta_{\rm ALMA}/2$---plausible, so we ascribe $\alpha_{\rm [NII]122, NLR}\le 0.2$, following the interpretation of F15. 
 
We use this upper limit to estimate, based on theoretical models, the corresponding [CII]158$\mu$m emission for given physical conditions prevalent in the NLR by writing
\begin{equation}
\alpha_{\rm [CII],NLR} = \frac{\gamma_{\rm [CII],NLR}^{\rm (G04)} \times F_{\rm [NII]122, NLR}}{F_{\rm [CII]}} \label{eq:alpha_cii_nlr_groves}
\end{equation}
where $F_{\rm [NII]122,NLR}$ ($=0.2 F_{\rm [NII]122}$) is the observed NLR flux of the [NII]122$\mu$m line based on the ALMA observation. The factor $\gamma_{\rm [CII],NLR}^{\rm (G04)}$ is, explicitly, written as
\begin{equation} \nonumber
\gamma_{\rm [CII],NLR}^{\rm (G04)} = \frac{F_{\rm [CII],NLR}^{\rm (G04)}}{F_{\rm [NII]122, NLR}^{\rm (G04)}},
\end{equation}
representing the scaling between the predicted fluxes of [CII]158$\mu$m and [NII]122$\mu$m---$F_{\rm [CII],NLR}^{\rm (G04)}$ and $F_{\rm [NII]122,NLR}^{\rm (G04)}$, respectively, from an NLR model of \citet[G04]{Groves2004}. We correct the nitrogen abundance from solar values adopted in G04 to match ISM values that we later use when considering the contribution to [CII]158$\mu$m from HII regions associated with star-formation; the ISM abundance set adopts a nitrogen abundance that is 1.3 times the solar nitrogen abundance.

Output of the G04 NLR models can be parametrized on grids of $n_{\rm H}$ and a dimensionless ionization parameter, $U = (\Phi_{\rm LyC})/(n_{\rm H} c)$, where $\Phi_{\rm LyC}$ is the rate of Lyman continuum photons per unit area from the AGN incident on the cloud surface, and $c$ is the speed of light. According to these grids, $\gamma_{\rm [CII],NLR}^{\rm (G04)}$ is relatively insensitive to $U$ and $n_{\rm H}$: $\gamma_{\rm [CII],NLR}^{\rm (G04)} =$ 4.6--7.7 throughout the parameter space for models with intrinsic power-law ionizing continua with spectral indices of -1.7 or -2.0, bounding the conditions for expected AGN flux. Adopting this range of $\gamma_{\rm [CII],NLR}$, we place an upper bound on the fraction of flux of [CII]158$\mu$m emerging from the compact nuclear NLR as $\alpha_{\rm [CII]158,NLR} \leq $ 0.15--0.25.

\subsubsection{\rm [CII]158$\mu\mathrm{m}$ from star-forming ISM} \label{subsec:pdrs_hii}

\paragraph{Ionized gas}

\begin{figure}[t]
\includegraphics[width=0.47\textwidth]{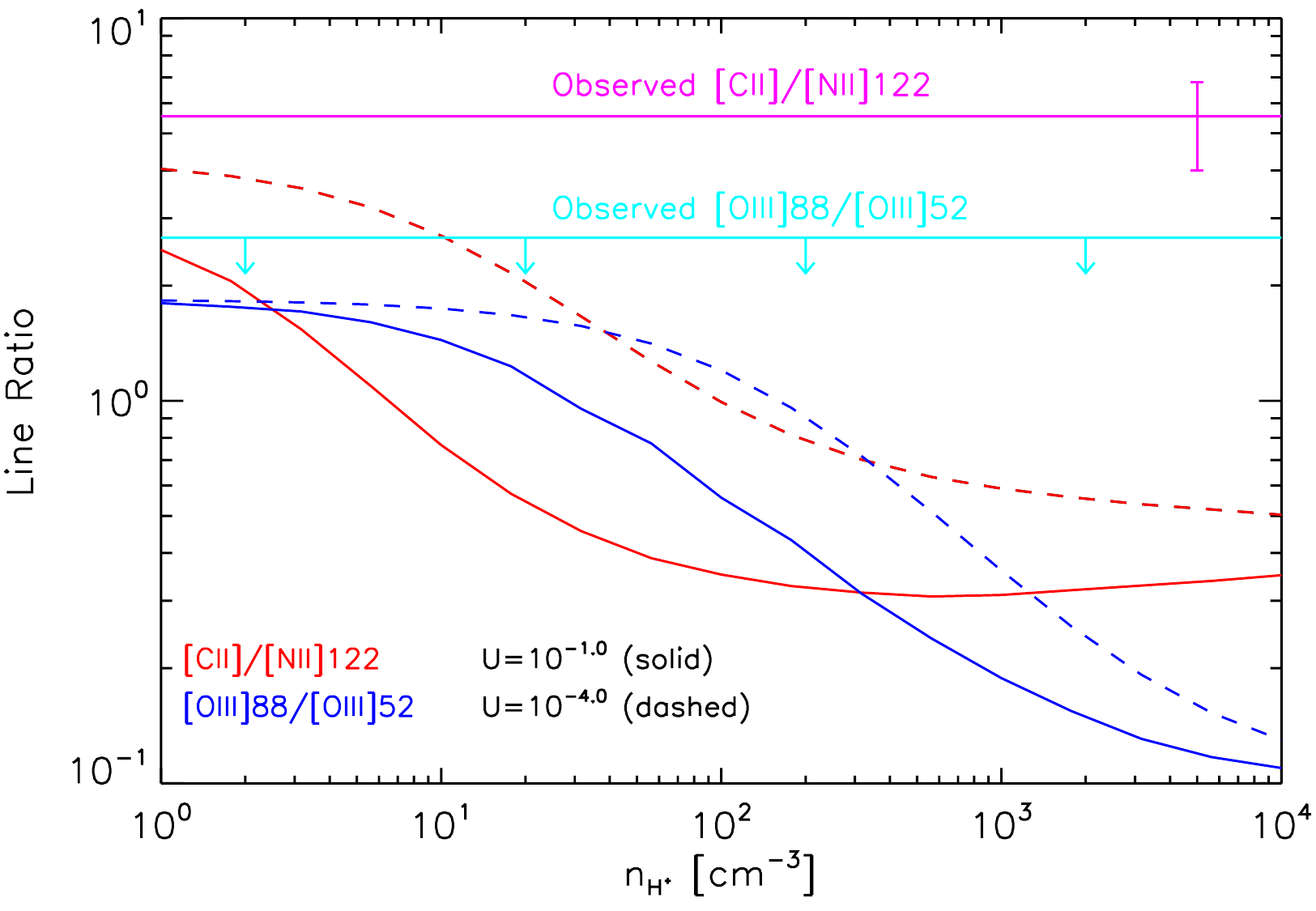}
\caption{Theoretical line flux ratios  $F_{\rm[CII],HII}^{\rm(\emph{Cloudy})}/F_{\rm[NII]122,HII}^{\rm(\emph{Cloudy})}$ and $F_{\rm[OIII]88,HII}^{\rm(\emph{Cloudy})}/F_{\rm[OIII]52,HII}^{\rm(\emph{Cloudy})}$  (red and blue curves, respectively) as a function of ionized gas density and computed for different ionization parameters ($U=-4.0$, dashed curves; $U=-1.0$, solid curves), computed for the HII region only. Magenta and cyan horizontal lines denote the measured value and upper limit of the respective ratios in the Cloverleaf. Measured fluxes of [CII]158$\mu$m and [NII]122$\mu$m have been corrected for NLR contributions according to $\alpha_{\rm [CII],NLR}\leq0.2$ and $\alpha_{\rm [NII]122,NLR}\leq0.2$, as discussed in Section~\ref{subsec:NLR}. Error bar indicates the range of uncertainty in the ratio after propagating uncertainties on the [CII]158$\mu$m and [NII]122$\mu$m fluxes. Downward pointing arrows indicate that the [OIII]88$\mu$m/[OIII]52$\mu$m flux ratio represents an upper limit, having used the $3\sigma$ upper limit reported for [OIII]88$\mu$m.}
\label{fig:ciinii}
\end{figure}

Turning now to the star-formation component in the Cloverleaf system, we first estimate the [CII]158$\mu$m emission from ionized gas. Here again we use the [NII] line measurements. With similar ionization potentials, singly-ionized nitrogen is often found in the same ionized gas as singly ionized carbon. Its slightly higher first ionization potential (IP = 14.5~eV), however, prevents the nitrogen ions from forming in neutral gas. Thus, identifying an excess in the measured flux ratio, $F_{\rm [CII]}/F_{\rm [NII]122}$, relative to a theoretical value, $F_{\rm [CII],HII}^{\rm(\emph{Cloudy})}/F_{\rm [NII]122,HII}^{\rm(\emph{Cloudy})}$, predicted for exclusively ionized gas, indicates the presence of additional [CII]-emitting components, such as PDRs. While the lower-level transition corresponding to the [NII]205$\mu$m line is better-suited to assess the fraction of [CII]158$\mu$m arising in ionized gas because $F_{\rm [CII]}/F_{\rm [NII]205}$ is density-independent, the [NII]122$\mu$m line can be used as well. We model its density-dependence with the photo-ionization code \emph{Cloudy}\footnote{We have used a plane-parallel geometry with ionizing spectrum from \emph{CoStar} stellar atmosphere model of \citet{Schaerer1997}.} \citep[version 10.0]{Ferland1998}; $F_{\rm [CII],HII}^{\rm(\emph{Cloudy})}/F_{\rm [NII]122,HII}^{\rm(\emph{Cloudy})}$ is shown in Figure~\ref{fig:ciinii} as functions of $U$ and $n_{\rm H^+}$, where $n_{\rm H^+}$ refers to the initial ionized gas density in our fixed-pressure HII region cloud model.

It is clear that the contribution of [CII]158$\mu$m from ionized gas can be large for low densities ($n_{\rm H^+}<10$ cm$^{-3}$) and low ionization parameters ($U\sim10^{-4}$). Note that the common ionized gas density probe, the [OIII]88$\mu$m-to-[OIII]52$\mu$m flux ratio $F_{\rm [OIII]88,HII}^{\rm(\emph{Cloudy})}/F_{\rm [OIII]52,HII}^{\rm (\emph{Cloudy})}$, is predicted to be less than 1.8 for all considered densities, $n_{\rm H^+}=1$--10$^4$~cm$^{-3}$. This value is below the ratio ($<2.3$) derived from the observed fluxes of these lines (Table~\ref{tab:flux_table}), which is not surprising as our measured ratio includes an upper limit on $F_{\rm [OIII]88}$. In this case, we must use an alternative means of estimating an average gas density in the Cloverleaf's star-forming, ionized ISM.

A detailed study of spatially resolved, large-scale emission from N$^{+}$ in the Galaxy indicates that observed emission from [NII]122$\mu$m and [NII]205$\mu$m arises from ionized gas characterized by $n_{\rm H^+}=10$--100~cm$^{-3}$ \citep{Goldsmith2015}, generally in agreement with previous observations \citep[e.g.,][]{Wright1991}, as well as a reflection of the fact that line emission tends to occur most efficiently at or near $n_{crit}$ ($\sim300$~cm$^{-3}$ for [NII]122$\mu$m). These inferred densities are up to several orders of magnitude higher than the diffuse ($n_{\rm H^{+}}=0.01$--0.1~cm$^{-3}$) warm ionized medium, but lower than densities found in the ionized bubbles of newly forming stars. Among the possible origins of such low density ionized gas mentioned in \citet{Goldsmith2015} are surfaces of dense molecular gas---copious in the Cloverleaf disk---irradiated by massive stars. Assuming then, that the average ionized gas density is the same as the average density derived in \citet{Goldsmith2015}: $n_{\rm H^+}= 30$~cm$^{-3}$. At this density, $F_{\rm[CII],HII}^{Cloudy}/F_{\rm[NII]122,HII}^{Cloudy}= 1.6$ for $U=10^{-4.0}$, and 0.46 for $U=10^{-1.0}$, according to Figure~\ref{fig:ciinii}. Comparing these values to our fiducial ratio, $F_{\rm [CII],HII}/F_{\rm [NII]122, HII} $ (with the NLR contributions removed):
\begin{equation}
\frac{F_{\rm [CII], HII}}{F_{\rm [NII]122, HII}} = \frac{F_{\rm [CII]}-F_{\rm [CII],NLR}}{F_{\rm [NII]122}-F_{\rm [NII]122,NLR}} = 5.5 \pm 1.7 \nonumber
\end{equation}
 (see Section~\ref{subsec:NLR}), suggests that $\alpha_{\rm [CII], HII}=0.2$. Here we have taken the average fraction inferred from the high ($\alpha_{\rm [CII], HII}=0.3$) and low ($\alpha_{\rm [CII], HII}=0.08$) ionization parameter cases to account for our uncertainty on $U$.

Conditions in the large-scale ISM of the Cloverleaf may be very different than those that \citet{Goldsmith2015} obtain in the Milky Way, due to the compactness of the emission and the presence of both a major starburst and a luminous AGN. However, observations of a broad range of sources, including star-forming regions, individual galaxies, and statistical samples of galaxies (cf. \citet{Oberst2006}, \citet{Rangwala2011}, and \citet{Vasta2010} for examples of each) indicate $\alpha_{\rm [CII],HII}<0.3$, which is consistent with our derived $\alpha_{\rm [CII],HII}$ using the Milky Way ionized gas density.

\begin{figure*}
\centering
\begin{tabular}{c c}
\includegraphics[width=0.374\linewidth]{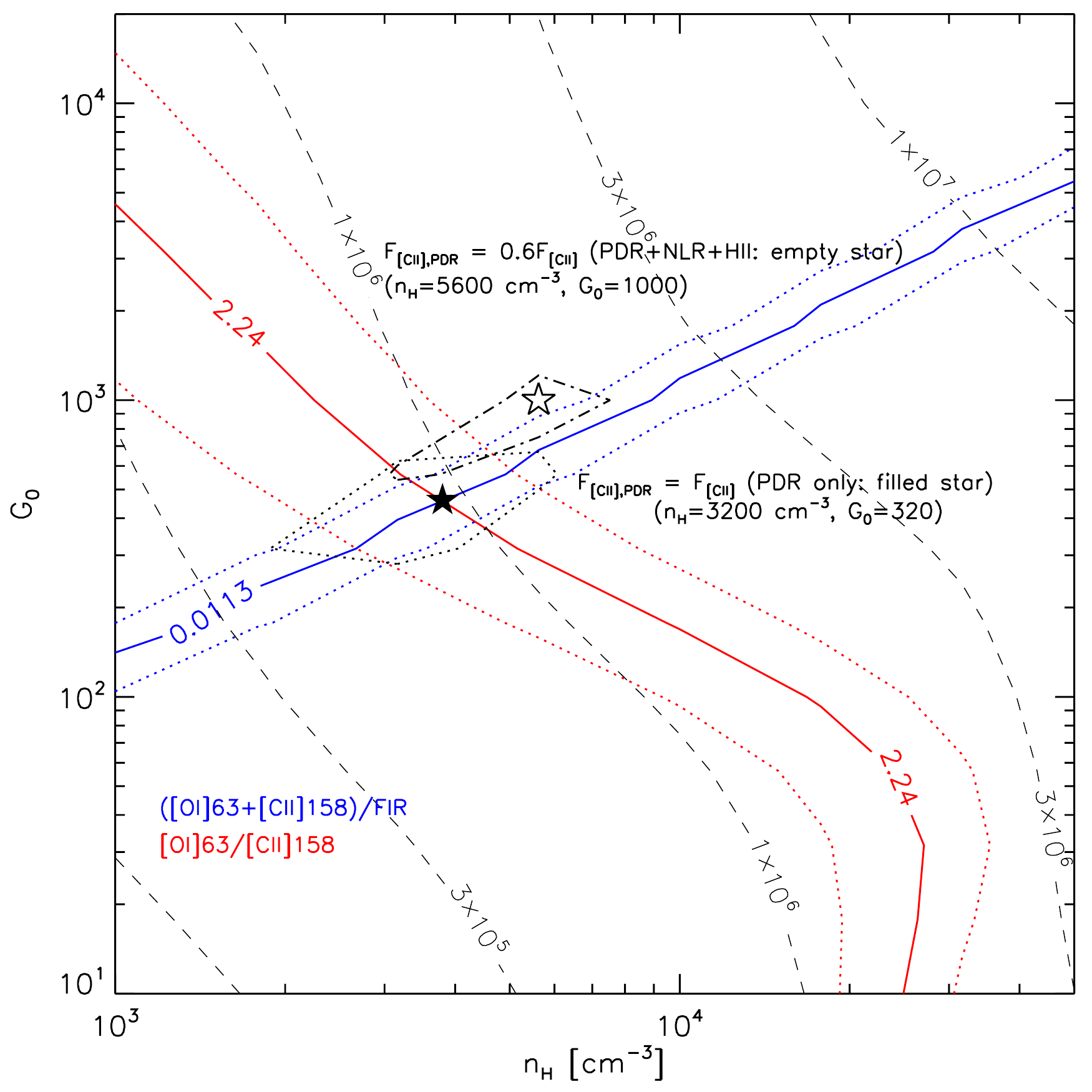} &
  \includegraphics[width=0.59\linewidth]{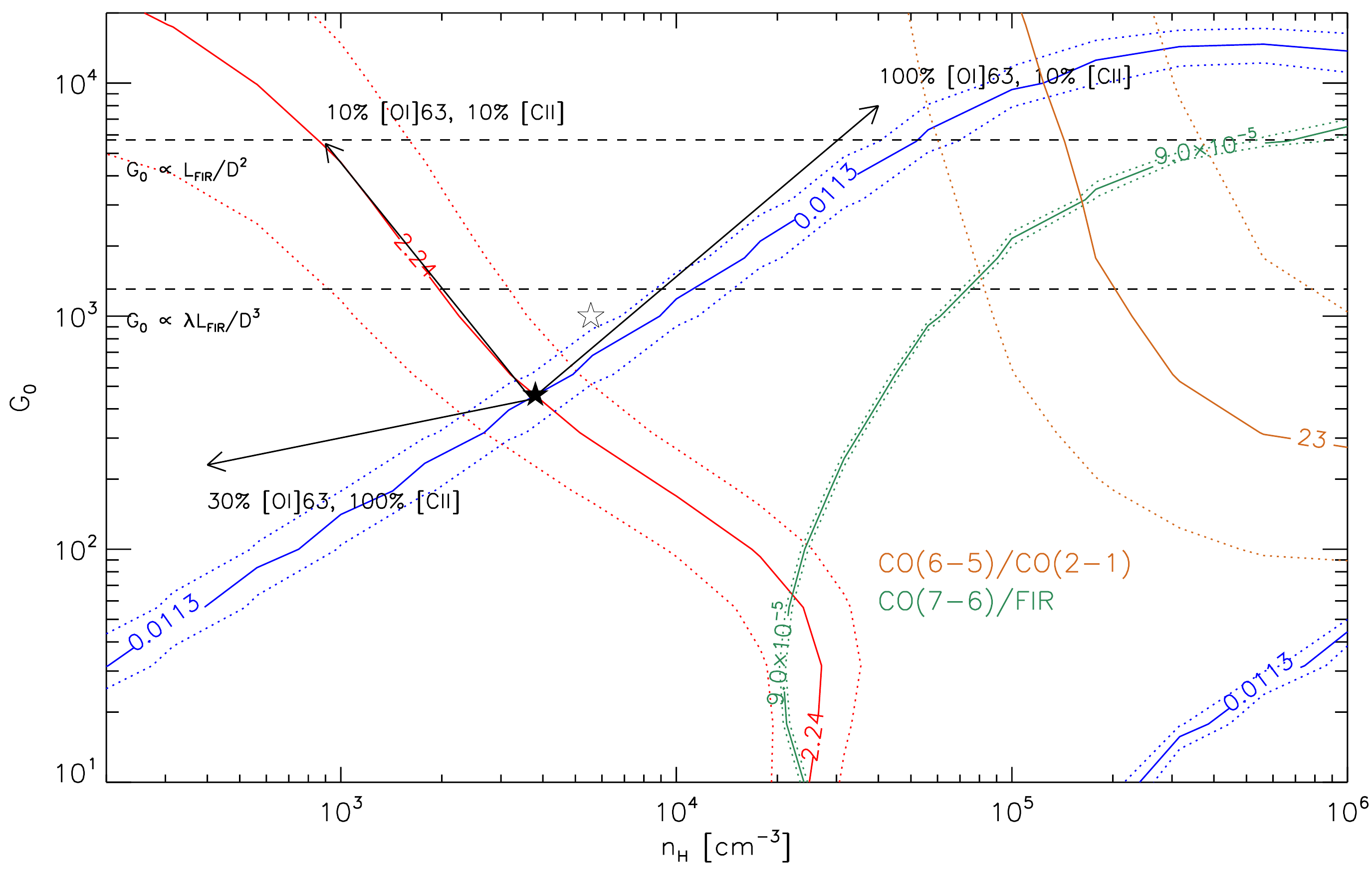} \\
\end{tabular}
\caption{PDR diagnostic plots. Left panel: Red and blue curves denote observed values (solid) and associated uncertainties (dotted) of flux ratios $F_{\rm [OI]}/F_{\rm [CII]}$ and $\left(F_{\rm [OI]}+F_{\rm [CII]}\right)/F_{\rm FIR}$, respectively. The filled black star symbol indicates the corresponding PDR solution in $n_{\rm H}$ and $G_0$, which refers to the PDR model with minimum $\chi^2$. The confidence region within one standard deviation from the PDR solution is outlined by the black dotted contour. The un-filled star represents the PDR solution with $F_{\rm [CII],PDR}=0.6F_{\rm[CII]}$, i.e., after applying corrections for NLR and HII region contributions to the measured [CII]158$\mu$m flux, and the black dot-dashed contour bounds the corresponding $1\sigma$ confidence region. Black dashed contours represent thermal pressure at the PDR surface. Right panel: Observed CO line flux ratios are shown along with the FS line ratios. The brown curve and purple curves show, respectively, $F_{\mathrm{CO}(J=7\rightarrow6)}/F_{\mathrm{CO}(J=2\rightarrow1)}$ and $F_{\mathrm{CO}(J=7\rightarrow6)}/F_{\rm FIR}$. Vectors signify the direction and magnitude of change in the PDR solution located at $n_{\rm H}=3,160$ cm$^{-3}$ and  $G_0=316$, when altering the PDR contribution to [OI]63$\mu$m and [CII]158$\mu$m flux. Percentages indicated by each vector refer to the percent of the FS lines originating from the PDR.}
\label{fig:K06diagnostic}
\end{figure*}

\paragraph{PDR diagnostics} Having subtracted the ionized gas contribution from both star-forming and non-star-forming ISM components to $F_{\rm [CII]}$, we are ready to incorporate the [CII]158$\mu$m and [OI]63$\mu$m fluxes, as well as the far-IR continuum, in the commonly-used framework of the PDR diagnostic diagram in Figure~\ref{fig:K06diagnostic}. We double the measured [OI]63$\mu$m flux, as per Kaufman et al. (1999), before comparing measurements to theoretical predictions from K06. This correction is necessary because, while the intensities computed for the PDR models emerge from a single, illuminated face, the geometry of emitting regions in the Cloverleaf is assumed to be such that individual PDRs are illuminated by FUV photons on all sides; optically thick [OI]63$\mu$m line emission emerging from cloud surfaces opposite the observer will not contribute to the measured flux.

According to Figure~\ref{fig:K06diagnostic} (lefthand panel), PDR models with $n_{\rm H}\sim3$--$7\times10^3$~cm$^{-3}$ and $G_0$\footnote{$G_0$ is normalized to the average value in the plane of the Milky Way, such that $G_0 =1$ indicates an FUV flux of $1.6\times10^{-3}$~erg~s$^{-1}$~cm$^{-2}$.} $\sim0.6$--$1.2\times10^3$ can generate line ratios that are compatible with observations interpreted in the context of our fiducial model, with $\alpha_{\rm [CII], NLR} = 0.2$ and $\alpha_{\rm [CII], HII}= 0.2$. Had we not made any allowances for possible contributions to $F_{\rm [CII]}$ from non-PDR gas, the K06 models would favor lower values of $n_{\rm H\sim2}$--$6\times10^3$ ~cm$^{-3}$ and $G_{0}\sim3$--$6\times10^2$. 

Note that we have made a couple implicit assumptions throughout this analysis, which we now address.

Firstly, we have assumed that all of the measured FIR luminosity is generated from the starburst component in the Cloverleaf, with negligible contamination from the AGN, as discussed in \citet{Lutz2007}. The notion that the FIR luminosity is dominated by star-formation in the host galaxy of a quasar is supported by recent PACS imaging of FIR emission in nearby quasar systems \citep{Lutz2016}. Uncertainties related to determining a precise contribution of the Cloverleaf's AGN to $\mathrm{L_{\rm FIR}}$ can be significant, however, and we note out that underestimating the AGN contribution leads to overestimating the derived $G_0$ and underestimating $n_{\rm H}$ by one order of magnitude each, and that no viable PDR solutions exist for cases where more than 60\% of the FIR luminosity is attributed to the AGN.

Secondly, we have assumed that the Cloverleaf NLR does not significantly contribute to the observed [OI]63$\mu$m flux, i.e., that the fraction of $F_{\rm [OI]}$ emitted from the NLR, $\alpha_{\rm [OI],NLR}$, is zero. Our fiducial PDR solution, however, is fairly robust to changes in $n_{\rm H}$ and $G_0$ as long as $\alpha_{\rm [OI],NLR}\le0.5$, so we do not undertake detailed partitioning of the [OI]63 flux, which depends strongly on the (currently poorly unconstrained) NLR density. Explicitly, we find the PDR solutions are $n_{\rm H}=5.6\times10^3$~cm$^{-3}$ and $G_0=1\times10^3$, $3.2\times10^3$~cm$^{-3}$ and $1\times10^3$, $1\times10^3$~cm$^{-3}$ and $1\times10^3$, for $\alpha_{\rm [OI],NLR}=0.0$--0.2, 0.3--0.4, and 0.5, respectively. According to the G04 NLR models considered in Section~\ref{subsec:NLR} could supply at most 50\% of $F_{\rm [OI]}$ in conditions where $n_{\rm H}\gtrsim5\times10^3$~cm$^{-3}$. We do not consider cases where $\alpha_{\rm [OI],NLR}>0.5$, because such large fractions would require appreciable dense NLR gas beyond the physical extent of the [NII]122$\mu$m region identified in F15.

\paragraph{PAH emission as a measure of $\alpha_{\rm [CII], PDR}$} The relatively constant ratio between [CII]158$\mu$m and total PAH emission observed in star-forming galaxies \citep{Helou2001, Croxall2012} across wide ranges of $L_{\rm IR}$ and dust temperatures reflects the fact that line emission from C$^+$ and photoelectron ejection from PAHs are dominant channels of gas cooling and heating, respectively, in PDRs \citep{Hollenbach1999}, and implies that we can use the previously observed PAH emission in the Cloverleaf to independently estimate $F_{\rm [CII], PDR}$. PAH emission features at 6.2$\mu$m and 7.7$\mu$m were detected in the Cloverleaf's rest-frame mid-IR spectrum using \emph{Spitzer}-IRS in \citet{Lutz2007}, and corresponding fluxes $F_{\rm PAH6.2}$ and $F_{\rm PAH7.7}$ were estimated by fitting Lorentzian profiles to the baseline-subtracted spectrum; no steps were taken for aperture correction as the Cloverleaf was unresolved in the IRS beam.

For the [CII]158$\mu$m-PAH relation, we use the analysis from D\'{i}az-Santos~et~al.~2016~(in preparation) of local (U)LIRGs in the Great Observatories All-sky LIRG Survey \citep[GOALS;][]{Armus2009}. The sample includes observations of [NII]205$\mu$m for more than 100 galaxies, which aids in the accurate partitioning of [CII]158$\mu$m between neutral and ionized gas phases \citep[e.g.,][]{Oberst2006}. Because flux estimation for the 7.7$\mu$m PAH feature is susceptible to greater uncertainties, due main to the difficulty in determining the underlying continuum in close proximity to the 9.7$\mu$m silicate absorption feature, we choose to use only the 6.2$\mu$m PAH feature for comparison between the Cloverleaf and GOALS galaxies.

The average [CII]158$\mu$m-to-PAH6.2$\mu$m flux ratio for the GOALS sample is $\sim0.3$, but there is appreciable scatter ($\sim0.3$~dex) with respect to the mean value, as well as a slight dependence on dust temperature, characterized by the ratio of continuum fluxes at 63~$\mu$m and 158~$\mu$m. Both effects can be interpreted as a result of considering only the [CII]158$\mu$m and PAH 6.2$\mu$m feature, rather than the respective sums of all major FS line emission and measurable fluxes in the PAH bands. We note, however, that the overall variation in the [CII]$158\mu$m-to-PAH6.2$\mu$m flux ratio with dust temperature is relatively small, and considerably less than the $\sim1$~dex variation in the [CII]158$\mu$m-to-FIR flux ratio with dust temperature as presented in \citet{DiazSantos2013}. To proceed with our comparison to GOALS galaxies, we measure the continuum flux densities $S_{63}=0.64 \pm 0.10$~Jy and $S_{158}=0.14 \pm 0.05$~Jy at 63~$\mu$m and 158~$\mu$m, respectively, for the Cloverleaf directly from the \emph{Herschel} spectra, and use the SED decomposition from \citet{Weiss2003} to remove the AGN contribution to $S_{63}$, finding $S_{63}/S_{158}=3.3$. With $F_{\rm [CII]}/F_{\rm PAH6.2} = 0.97\pm0.26$ (where the uncertainty only reflects uncertainty in $F_{\rm [CII]}$), this places the Cloverleaf system within the range of dust temperatures and [CII]158$\mu$m (total)-to-PAH6.2$\mu$m observed in GOALS, although at the high end of values spanned by both quantities. Relative to GOALS galaxies with comparable dust temperature, which have [CII]158$\mu$m (PDR-only)-to-PAH6.2$\mu$m ratios $\sim0.2$--0.3, the high value of $F_{\rm [CII]}/F_{\rm PAH6.2}$ in the Cloverleaf suggests a large contribution of  $\sim70$\% to $F_{\rm [CII]}$ from non-PDR gas. A number of uncertainties exist when using this method to estimate $\alpha_{\rm [CII], PDR}$, however, including those related to (1) the SED fitting, (2) extinction in the PAH 6.2$\mu$m feature, and (3) PAH flux estimation. We consider (1) and (3) as subject to the largest sources of uncertainty. The SED decomposition is sensitive to various assumptions regarding, for example, dust opacity, and a larger AGN contribution to $S_{63}$ or a cooler dust temperature as found in F15 for the starburst component would suggest a larger PDR fraction of $F_{\rm [CII]}$. Also, $F_{\rm PAH6.2}$ strongly depends on different systematics of the adopted line fitting procedures for the GOALS analysis and for \citet{Lutz2007}, and is likely underestimated in the Lorentzian fit, which does not take into account the wings of the PAH feature. As in the case of a cooler starburst temperature, a larger PAH6.2$\mu$m flux in the Cloverleaf would increase the PDR fraction inferred from the GOALS relation. We therefore consider the 30\% contribution determined here to be a lower limit on $\alpha_{\rm [CII],PDR}$, although still within a factor of $\sim2$ of the true value.

\subsubsection{Atomic mass estimate} The optically thin [CII]158$\mu$m emission can be used to assess the mass of [CII]158$\mu$m-emitting atomic gas, $M_{\rm H}$, by comparing the [CII]158$\mu$m luminosity attributed to PDRs, $L_{\rm [CII], PDR}$, to the expected C$^+$ cooling rate (units of erg~s$^{-1}$ per H atom). Following \citet{HaileyDunsheath2010} (cf. their Equation 1), and using the PDR surface temperature $T_{\rm PDR}\sim300$~K and density $n_{\rm H}=5.6\times10^3$~cm$^{-3}$---the fiducial PDR solution discussed above---we find $M_{\rm H} \sim 1.8\times 10^{10}$~M$_{\odot}$. While there is a large (factor of $\sim2$) uncertainty on the mass estimate, this value is on the high end of the Cloverleaf's molecular gas mass estimated by B09 ($M_{\rm H_2}\sim$ 0.3--3$\times10^{10}$~M$_{\odot}$), indicating that the mass in PDRs is comparable to the molecular mass in H$_2$.

\subsubsection{Geometric constraints on the far-UV field} \label{subsubsec:geo_gnot}
Following \citet{Stacey2010} and authors thereafter, it is useful to compare the value of $G_0$ predicted from the PDR models with the value estimated solely from geometric considerations. As described in \citet{Wolfire1990}, if PDR surfaces and the sources of radiation are randomly distributed---such as in the case of stellar populations---within a region of diameter $D$, then the photons in this region will likely be absorbed by a PDR cloud before traveling a distance $D$ so that the incident FUV flux on cloud surfaces in the emitting region is simply related to the surrounding volume density of FUV photons, $G_0\propto (\lambda L_{\mathrm{FIR}})/D^3$, where $\lambda$ is the mean-free path of FUV photons. However, if FUV photons can travel large distances (compared to $D$) until being absorbed, then the FUV flux will vary as the surface flux of FUV photons, $G_0\propto L_{\rm FIR}/D^2$. The constants of proportionality can be determined by calibrating to known values\footnote{For M82, $G_0 = 10^{2.8}$ and $L_{\mathrm{FIR}} = 3.2\times10^{10}$ \citep{Colbert1999}, and $D=300$ pc \citep{Joy1987}} of $G_0$, $L_{\mathrm{FIR}}$, and $D$ for M82 (as in, for example, \citet{Stacey2010}). These two scenarios represent limiting cases for which we can calculate the expected $G_0$, given $D$ and $L_{\rm FIR}$ for the Cloverleaf. 

To make this comparison, we adopt the source size $D=1.3$~kpc inferred from gravitational lens modeling (VS03) of spatially resolved CO(7-6) flux, implicitly assuming that the CO(7-6) line emission is co-spatial with the FIR continuum. For the Cloverleaf, there exists spatially resolved continuum data (F15) in the rest-frame FIR (122$\mu$m) that would---barring significant contribution at this wavelength from the AGN as inferred from the double-peaked SED in \citet{Weiss2003}---enable a direct measurement of the extent of the FIR-emitting region in the context of a gravitational lens model that relates the observed surface brightness distribution to a physical size in the source plane. A comparison between the 122$\mu$m continuum and CO(7-6) maps shows that the two tracers peak at the same locations in the four lensed quasar images, supporting our claim that the two emission regions overlap. Setting $D=1.3$~kpc, then, in the analytic expressions for $G_0$ when $\lambda\gg D $ and $\lambda \ll D$, we estimate $G_0\sim$ 1.3--5.7$\times10^3$.

This inferred $G_0$ is above the value predicted by the fiducial PDR model in Section~\ref{subsec:pdrs_hii}, which points to $G_0=1\times10^3$. Referring to Figure~\ref{fig:K06diagnostic} (righthand panel), which shows the effects of varying $\alpha_{\rm [CII], PDR}$ and $\alpha_{\rm [OI], PDR}$ on derived PDR parameters, we see that higher values of $G_0$ are suggestive of larger contributions from non-PDR gas to both $F_{\rm [OI]}$ and $F_{\rm [CII]}$, or to $F_{\rm [CII]}$ only. Given the uncertainty associated with the analytic estimates, we do not find this to be strong evidence in favor of additional ISM components unexplored up to this point in the analysis, concluding instead that the fiducial PDR solution broadly agrees with the analytic estimate. However, following a different line of reasoning, we will consider additional components, namely, X-ray heated gas, in Section~\ref{subsec:XDR}.

\subsection{CO from PDR models} Next we compare the PDR diagnostic diagram constructed with FS line ratios to previous attempts at interpreting CO emission in the Cloverleaf using the PDR paradigm. In B09, the authors found that line ratios from K06 PDR models with densities and FUV fields in the range of $n_{\rm H}=$ 1--4$\times10^5$~cm$^{-3}$ and $G_0=$1--5$\times10^3$ provided a suitable match to the measured CO line ratios $F_{\mathrm{CO}(J=7\rightarrow6)}$/$F_{\rm FIR}$ and $F_{\mathrm{CO}(J=6\rightarrow5)}$/$F_{\mathrm{CO}(J=2\rightarrow1)}$. ($F_{\mathrm{CO}(J=2\rightarrow1)}$ used here is the flux expected if the line is thermalized at the same temperature as the observed CO($J=3\rightarrow2$) line.)   In the righthand panel of Figure~\ref{fig:K06diagnostic}, we supplement the analysis in B09 with the observed, extinction-corrected [OI]63$\mu$m and [CII]158$\mu$m fluxes. For consistency, we show the same pair of CO diagnostic ratios used in B09, and find that the FS line ratio diagnostics favor lower $n_{\rm H}$ and $G_0$ than the CO diagnostics. Notably, if the origin of observed CO transitions could indeed be traced to PDRs with densities of order $\sim10^5$ cm$^{-3}$, we would have expected this emission to be accompanied by much higher [OI]63$\mu$m flux than is observed. The high density solution inferred from these mid-J CO line ratios overproduces [OI]63$\mu$m by factors of 3--20 for the range of PDR solutions identified in B09; or, alternatively, the low density solution suggested by the FS lines under-predicts the observed $F_{\mathrm{CO}(J=7\rightarrow6)}$ by factors of 10--100. For illustrative purposes, we have drawn vectors which show the direction and magnitude of change in $n_{\rm H}$ and $G_0$, corresponding to different assumptions of $\alpha_{\rm [CII], PDR}$ and $\alpha_{\rm [OI], PDR}$. \emph{Importantly, we find that there exists no combination of $\alpha_{\rm [CII], PDR}$ and $\alpha_{\rm [OI], PDR}$ that can eliminate the discrepancy in mid-J CO- and FS line-derived PDR conditions, leading us to conclude that PDR conditions inferred from [OI]63$\mu$m and [CII]158$\mu$m fluxes fail to describe the observed mid-J CO emission.} Similarly, increases in the AGN contribution to $L_{\rm FIR}$ and differential lensing of emission lines do not lead to consistent PDR solutions between the FS- and CO-based diagnostics.
 
\begin{figure}
\includegraphics[width=0.47\textwidth]{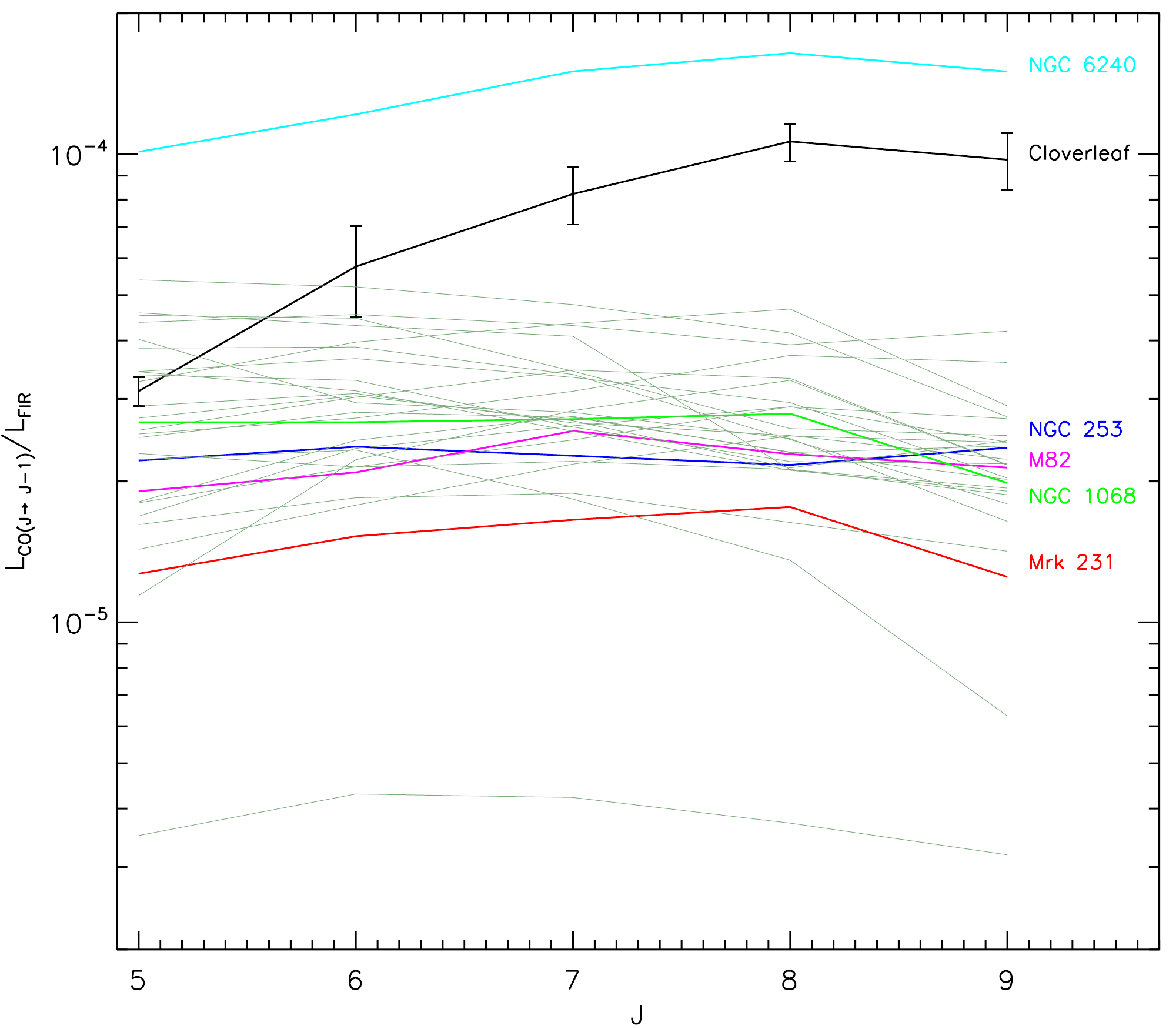}
\caption{Cloverleaf CO spectral line energy distribution (SLED), normalized to $L_{\rm FIR}$, compared to local (U)LIRGs (grey curves) of the HerCULES sample \citep{Rosenberg2015}. CO SLEDs for HerCULES galaxies NGC~253, NGC~1068, NGC~6240, and Mrk~231 are highlighted in blue, green, red, and cyan, respectively. M~82 (magenta curve) is also included here with data from \citet{Panuzzo2010}.}
\label{fig:coSLEDs}
\end{figure}

Furthermore, the high fluxes of the mid-J CO transitions relative to the FIR continuum flux observed in the Cloverleaf were noted in B09 as being anomalously high compared to typical fluxes observed in nearby ($z\sim0$) starbursts. In Figure~\ref{fig:coSLEDs}, we show a comparison to 27 local Luminous Infrared Galaxies (LIRGs; $L_{\rm IR} \geq 10^{11}$ L$_{\odot}$) and Ultra Luminous Infrared Galaxies (ULIRGs; $L_{\rm IR}>10^{12}$ L$_{\odot}$) in the \emph{Herschel} Comprehensive ULIRG Emission Survey (HerCULES) sample, which is a subset of the GOALS sample \citep{Rosenberg2015}\footnote{The full HerCULES sample includes 29 galaxies. We have excluded 2 galaxies with spectra containing [OI]63$\mu$m in absorption.}. It is clear that the Cloverleaf exhibits both unusually high CO line-to-FIR continuum luminosity ratios for transitions $J>6$, and a higher level of excitation in its spectral line energy distribution (SLED) than most of the galaxies---including 12 starbursts, 3 AGN, and 11 composite starburst/AGN systems. A notable exception is the known merger remnant and starburst/AGN composite system, NGC~6240. Quantitatively, the sum of FIR-normalized CO luminosities from $J=5$ to $J=9$, $\sum_{J=5}^{J=9} L_{\mathrm{CO}(J\rightarrow J-1)}/L_{\rm FIR}$, is, on average, equal to $1.5\times10^{-4}$ for the HerCULES galaxies, and $3.7\times10^{-4}$ for the Cloverleaf. We reach the same conclusion when comparing the Cloverleaf's CO emission to a more comprehensive compilation of CO SLEDs in \citet{Greve2014}, which includes 20 purely star-forming galaxies from the HerCULES sample, as well as additional star-forming local (U)LIRGs, and a statistically significant sample of high redshift ($0.1<z<6$) starbursts. Authors in both \citet{Greve2014} and \citet{Rosenberg2015} similarly disfavor far-UV heating from SF as the sole excitation mechanism for CO in the majority of the galaxies studied, given their elevated CO-to-FIR ratios and high excitation levels compared to the Milky Way or M~82, citing mechanical heating deposited to the gas from supernovae-driven turbulence or cosmic-rays, for example, as possible alternatives in sources with little or no contribution from AGN. We consider these possibilities for the Cloverleaf in Section~\ref{sec:mech_heating}.

\subsection{X-ray dominated regions} \label{subsec:XDR}

X-ray energy released from an accreting SMBH was proposed in B09 as an alternate source of  gas heating responsible for the observed level of CO emission, since the close proximity of dense molecular gas in the Cloverleaf disk to the central AGN could result in conditions favorable to the formation of X-ray Dominated Regions (XDRs). XDRs can produce strong emission, as in PDRs, from O$^0$, C$^+$, and CO, because the attenuation of X-ray flux is much slower in XDRs and the gas heating more efficient, thus often resulting in higher emergent intensities of these spectral lines from XDR cloud surfaces and larger line-to-$L_{\rm FIR}$ ratios. The relative importance of each line transition in the gas cooling depends mainly on the X-ray flux and gas density.
 
Considering only the total CO cooling, B09 identified a region in the parameter space of radial distance, $R$, from the AGN and gas density, namely, $500 \mathrm{\ pc}<R<1500\ \mathrm{pc}$ and $n_{\rm H}>3\times10^4$ cm$^{-3}$ for which the resulting emergent CO surface flux\footnote{Throughout this paper, we refer to ``surface flux" (in units of erg~s$^{-1}$~cm$^{-2}$) as the distribution of the emitted luminosity over a given surface area, or, equivalently, surface brightness integrated (in units of erg~s$^{-1}$~cm$^{-2}$~sr$^{-1}$) over a solid angle of 2$\boldsymbol{\pi}$.} from corresponding XDR models, $S_{\rm CO,XDR}^{\rm (B09)}$ (units of erg~s$^{-1}$~cm$^{-2}$), matched the observed CO surface flux, $S_{\rm CO}$, in the Cloverleaf disk for a range of areal filling factors, $\phi_{\rm ff}$. Note that radial distance is a proxy for the X-ray surface flux $S_{\rm X}$\footnote{for a surface defined at $R$}, when assuming, as in B09, fixed values of the hard (2--10 keV) X-ray luminosity, $L_{\rm X}$, and attenuating column of foreground hydrogen, $N_{\rm H_{\mathrm{att}}}$. For reference, an X-ray luminosity of $10^{46}$ erg s$^{-1}$ and attenuating hydrogen column of $3\times10^{23}$ cm$^{-2}$ translates to an X-ray surface flux of 10~erg~s$^{-1}$~cm$^{-2}$ at $R = 600$~pc. We note that in this framework, a larger distance is equivalent to a model with a larger attenuating column and/or lower $L_{\rm X}$ than the fiducial model considered here.

In this section, we examine whether the predicted [CII]158$\mu$m and [OI]63$\mu$m surface fluxes, $S_{\rm [CII],XDR}^{\rm (B09)}$ and $S_{\rm [OI],XDR}^{\rm (B09)}$, associated with the corresponding CO-constrained XDR models from B09 are consistent with observations. Furthermore, we determine the relative importance of X-ray versus stellar UV heating in producing the observed FS line cooling. We make use of the same grid of theoretical XDR models presented in B09, which are based on improved models of \citet{Maloney1996}, to determine $\alpha_{i, \rm XDR}$. If we express $S_{\rm [CII],XDR}^{\rm (B09)}$ and $S_{\rm [OI],XDR}^{\rm (B09)}$ in terms of the observed XDR flux of CO, $F_{\rm CO,XDR}$, then we can write the respective fractions $\alpha_{\rm [OI], XDR}$ and $\alpha_{\rm [CII], XDR}$ as 
\begin{align}
\alpha_{\rm [OI],XDR} &= \frac{\gamma_{\rm [OI],XDR}^{\rm(B09)} \times F_{\rm CO, XDR}}{F_{\rm [OI]}} \label{eq:alpha_oi63_xdr} \\
\alpha_{\rm [CII],XDR} &= \frac{\gamma_{\rm [CII],XDR}^{\rm(B09)} \times F_{\rm CO,XDR}}{F_{\rm [CII]}}, \label{eq:alpha_cii_xdr}
\end{align}
where 
\begin{align}\nonumber
\gamma_{\rm [OI],XDR}^{\rm(B09)} &= \frac{S_{\rm [OI],XDR}^{\rm(B09)}}{S_{\rm CO,XDR}^{\rm (B09)}} \\
\gamma_{\rm [CII],XDR}^{\rm(B09)} &= \frac{S_{\rm [CII],XDR}^{\rm(B09)}}{S_{\rm CO,XDR}^{\rm(B09)}} \nonumber
\end{align}
Equations~\ref{eq:alpha_oi63_xdr} and~\ref{eq:alpha_cii_xdr} implicitly assume that the [CII]158$\mu$m, [OI]63$\mu$m, and CO emission arise from the same XDR component characterized by a single areal filling factor $\phi_{\rm ff}$.   

The precise value of $F_{\rm CO,XDR}$ depends on what fraction of $F_{\rm CO}$ is attributed to the XDR. While B09 found successful XDR models assuming $F_{\rm CO,XDR}= F_{\rm CO}$, here we include the possibility of a star-formation component for the CO emission. Thus, we consider two limiting cases where 1) the starburst contribution to CO is minimal and we can attribute all of the observed CO to an XDR such that $F_{\rm CO,XDR} = F_{\rm CO}$ and 2) the starburst contribution to CO is ``M~82-like" and amounts to $2\times10^{-4}$ of the observed FIR luminosity, as per M~82 \citep{Kamenetzky2014}, leading to $F_{\rm CO,XDR}= 0.7F_{\rm CO}$. Note that M~82 is a good reference source given the similarity of its SLED to the sample of CO SLEDs in local (U)LIRGs displayed in Figure~\ref{fig:coSLEDs}. 
\begin{figure*}[t]
\centering
\begin{tabular}{c c}
\includegraphics[width=0.48\linewidth]{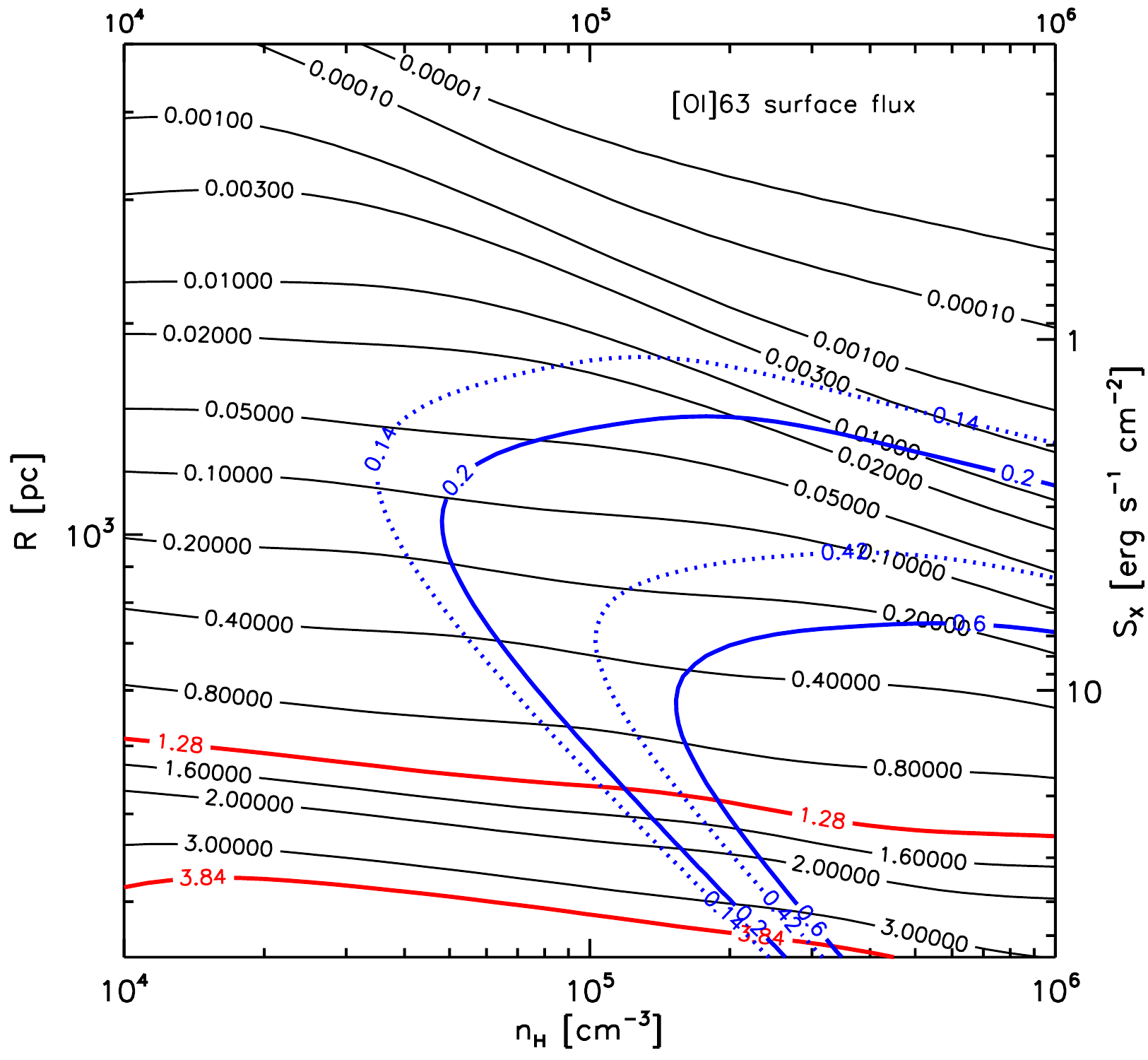} &
\includegraphics[width=0.48\linewidth]{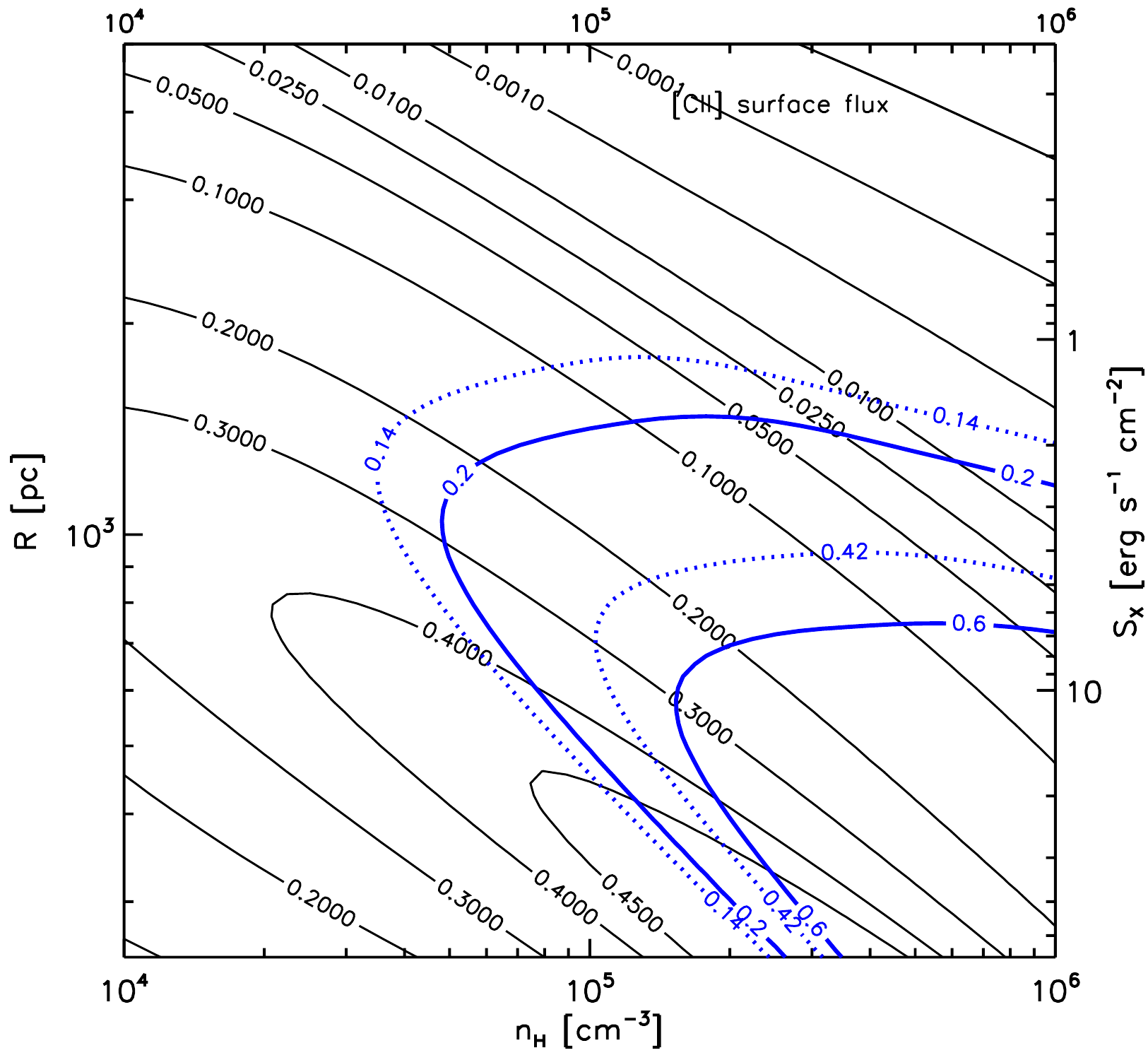} \\
\end{tabular}
\caption{XDR predictions for [OI]63$\mu$m (left panel) and [CII]158$\mu$m (right panel) surface flux, computed as a function of distance, $R$, from the AGN and density, $n_{\rm H}$ for a model grid with $N_{\rm H,att}=3\times10^{23}$ cm$^{-2}$ and $L_{\rm X}=10^{46}$ erg s$^{-2}$. Thick magenta curves indicate observed FS line surface flux for [OI]63$\mu$m, assuming area filling factors of 1 and $1/3$. Thick blue curves denote CO surface flux for the case $S_{\rm CO, XDR}=S_{\rm CO}$, with filling factors of unity and 1/3. Thin dotted blue contours represent the case where $S_{\rm CO, XDR}=0.7 S_{\rm CO}$ for the same two filling factors.} 
\label{fig:XDRoi63cii}
\end{figure*}

Figure~\ref{fig:XDRoi63cii} shows the expected surface \text{flux} (black contours) for [OI]63$\mu$m and [CII]158$\mu$m, denoted $S_{\rm [OI],XDR}^{\rm(B09)}$ and $S_{\rm [CII],XDR}^{\rm(B09)}$, produced in the B09 XDR model grid as functions of radius and density. Contours for $S_{\rm CO,XDR}^{\rm(B09)}$ indicating 0.6 and 0.2~erg~s$^{-1}$~cm$^{-2}$ are redrawn here from B09 (cf. their Figure 7) to facilitate comparison with results for the FS lines. These values represent partial ($\phi_{\rm ff}=1/3$) and complete ($\phi_{\rm ff}=1$) areal coverage, respectively, of the observed total CO luminosity ($=3.3\times10^9$ L$_{\odot}$) distributed over the surface of the disk, and bound the allowed parameter space of $R$ and $n_{\rm H}$ for the XDR models which successfully describe the observed CO emission in the case where $F_{\rm CO,XDR} = F_{\rm CO}$. Also shown are a pair of surface flux contours ($S_{\rm CO,XDR}^{\rm (B09)}$ = 0.42 and 0.14~erg~s$^{-1}$~cm$^{-2}$) for the same area filling factors $\phi_{\rm ff}=1/3$ and 1, computed for the case where $F_{\rm CO,XDR} = 0.7F_{\rm CO}$, described above. Thick red contours on the [OI]63$\mu$m surface flux map indicate the observed surface flux for [OI]63$\mu$m for $\phi_{\rm ff}=1$ and 1/3, namely, 1.28 and 3.84 erg s$^{-1}$ cm$^{-2}$, respectively; the [CII]158$\mu$m surface flux predicted from the B09 XDR model never reaches the observed values for the filling factors considered. 

For [OI]63$\mu$m, models at radii less than the VS03 disk radius of 650~pc predict a line surface flux comparable to or larger than the corresponding CO surface flux. At the smallest radii in the model grid, it can be as much as $\sim$6--20 times brighter than the observed CO, depending on filling factor: $S_{\rm [OI],XDR}^{\rm (B09)}$ and $S_{\rm CO,XDR}^{\rm (B09)}$ are roughly orthogonal in this regime, so we can keep the predicted CO surface flux fixed while rapidly increasing the [OI]63$\mu$m surface flux. These large [OI]63$\mu$m surface fluxes, when converted to luminosities via their respective area filling factors, amount to luminosities that are greater than the measured $L_{\rm [OI]}$, i.e., correspond to models with  $\alpha_{\rm [OI],XDR}>1$. We can then use the observed [OI]63 luminosity to identify a minimum distance, $R_{min}$, for XDR models that satisfy constraints by both FS line and CO emission. This bound occurs at higher radii ($R_{min}=450$ pc) for models with $\phi_{\rm ff}$ close to unity than for models with smaller values of $\phi_{\rm ff}$, which begin to produce unrealistically high [OI]63$\mu$m luminosities at $R_{min}\le250$ pc. The choice of a minimal or M82-like starburst paradigm has negligible effect on deriving $R_{min}$, as the contours of $S_{\rm CO,XDR}^{\rm(B09)}$ are very tightly spaced at $R<650$ pc. 

At larger radii, $S_{\rm [OI],XDR}^{\rm(B09)}$ can be less than the observed CO surface flux, with the XDR models contributing at most 30\% of the measured [OI]63$\mu$m flux at $R>650$~pc. At such radii and high densities, $S_{\rm [OI], XDR}^{\rm(B09)}$ is $\sim 2$ orders of magnitude below $S_{\rm CO, XDR}^{\rm(B09)}$, and $\alpha_{\rm [OI],XDR}$ is of order $\sim1$\% in this regime. 

For [CII]158$\mu$m, the observed surface flux in the Cloverleaf is between $1.15$~erg~s$^{-1}$~cm$^{-2}$ and $3.45$~erg~s$^{-1}$~cm$^{-2}$ for $\phi_{\rm ff}$ between 1 and 1/3. Unless $\phi_{\rm ff}>1$, there are no XDR models tested which can account for all of the measured luminosity, which agrees with the expectation that [CII]158$\mu$m traces primarily the star-formation process. At most, for each assumption about the starburst contribution to CO luminosity, $S_{\rm [CII],XDR}^{\rm(B09)}$ can be twice and three times as much, respectively, as $S_{\rm CO,XDR}^{\rm(B09)}$ in the acceptable parameter space at $R <650$~pc and $n_{\rm H}>3\times10^4$~cm$^{-3}$. Like $S_{\rm [OI],XDR}^{\rm (B09)}$, $S_{\rm [CII],XDR}^{\rm (B09)}$ is small compared to $S_{\rm CO,XDR}^{\rm(B09)}$ for large radii ($R\gtrsim1000$~pc) and high densities ($n_{\rm H}\gtrsim10^5$~cm$^{-3}$), and $\alpha_{\rm [CII],XDR}\sim1\%$ here. 

\begin{figure}
\centering
\includegraphics[width=0.44 \textwidth]{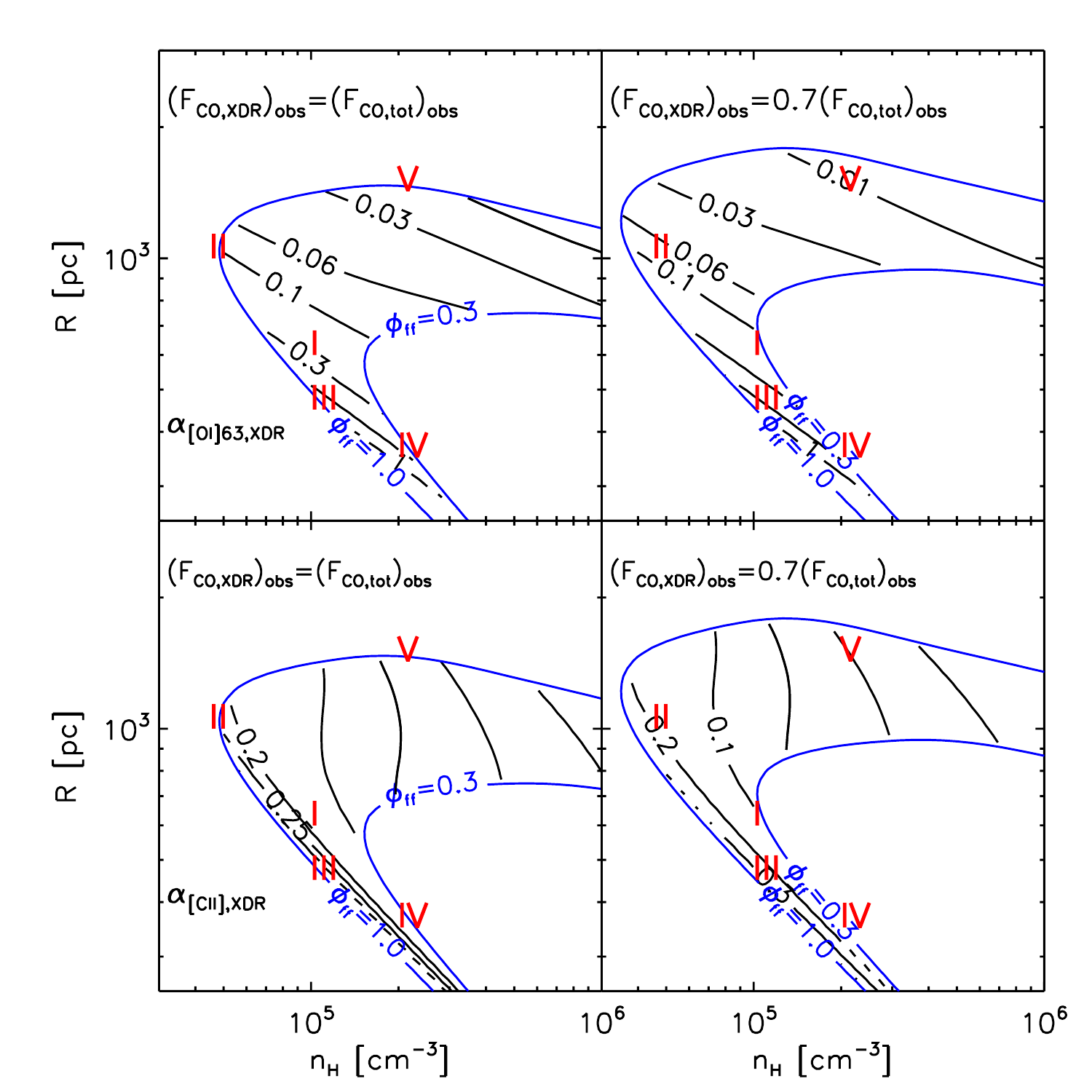} \\
\caption{Contours of constant $\alpha_{\rm [OI], XDR}$ (top panel) and $\alpha_{\rm [CII], XDR}$ for XDR models that reproduce the observed CO surface flux with area filling factors between $\phi_{\rm ff}=0.3$ and 1, according to the minimal and M~82-like starburst paradigms. Roman numerals ``I", ``II", ``III", ``IV", and ``V" highlight positions in the $R$-$n_{\rm H}$ plane that we consider as representative cases of various qualitatively distinct regimes in the allowed parameter space.} 
\label{fig:alphaoi63ciiXDR}
\end{figure}

It is clear from Figure~\ref{fig:XDRoi63cii} that $\alpha_{i,\rm XDR}$ varies widely throughout the allowable parameter space. Figure~\ref{fig:alphaoi63ciiXDR} depicts $\alpha_{\rm [OI], XDR}$ and $\alpha_{\rm [CII], XDR}$ as a functions of $R$ and $n_{\rm H}$ for the CO-constrained models with $1<\phi_{\rm ff} < 1/3$. (Recall that, for the minimal and M~82-like starbursts, respectively, area filling factors $\phi_{\rm ff}=$ 1--$1/3$ correspond to $S_{\rm CO,XDR}^{\rm (B09)} =$ 0.2--0.6~erg~s$^{-1}$~cm$^{-2}$ and 0.14--0.42~erg~s$^{-1}$~cm$^{-2}$.) In this figure, $\alpha_{i,\rm XDR}$ has been computed at each locus in $R$ and $n_{\rm H}$ according to Equations~\ref{eq:alpha_oi63_xdr} and ~\ref{eq:alpha_cii_xdr}. Again, we have assumed identical filling factors for CO and FS line emission.

Lacking additional constraints to guide us to fiducial values of $R$ and $n_{\rm H}$, we partition the available parameter space of the XDR model grid into qualitatively distinct regions of $R$ and $n_{\rm H}$ to provide a comprehensive summary of expected contributions to the FS lines from the CO-constrained XDR. The various regimes considered are:
\begin{enumerate}[I:]
\item Moderate radius/moderate density ($R=600$ pc and $n_{\rm H} = 10^5$~cm$^{-3}$)
\item Large radius/low density ($R=1000$ pc and $n_{\rm H} = 5\times10^4$~cm$^{-3}$)
\item Small radius/moderate density ($R = 500$ pc and $n_{\rm H}=10^5$~cm$^{-3}$)
\item Small radius/high density ($R=350$ pc and $n_{\rm H}=2\times10^5$~cm$^{-3}$)
\item Large radius/high density ($R = 1430$ pc and $n_{\rm H} = 2\times10^5$~cm$^{-3}$)
\end{enumerate}
Each case is marked on Figure~\ref{fig:alphaoi63ciiXDR}, and is associated with a particular value of $\alpha_{\rm [OI], XDR}$ and $\alpha_{\rm [CII], XDR}$, which is recorded in Table~\ref{tab:alpha_ism}. Broadly, we find that the CO-constrained XDRs are capable of producing between 20--40\% of observed [CII]158$\mu$m and 5--20\% of [OI]63$\mu$m for moderate X-ray surface fluxes, $S_{\rm X}$ of order $\sim1.6$--10~erg~s$^{-1}$~cm$^{-2}$, and densities $n_{\rm H}=$ 0.5--1$\times10^5$~cm$^{-3}$ (Cases I and II). For higher $S_{\rm X}$ and $n_{\rm H}$, namely, $S_{\rm X}\sim$ 10--30~erg~s$^{-1}$~cm$^{-2}$, $n_{\rm H}=$ 1--2$\times10^5$~cm$^{-3}$, the contributions increase for both FS lines, and X-ray heating can provide the main source of gas heating for the observed [OI]63$\mu$m emission: $\alpha_{\rm [CII],XDR}=$ 0.2--0.4 and $\alpha_{\rm [OI]63, XDR} =$ 0.7--0.8 (Case III and IV). For the case of high densities and low fluxes, however, there is negligible ($<5\%$) contribution to the FS lines (Case V), and stellar UV heating in PDRs clearly dominates as the mechanism responsible for producing the observed FS line cooling in this regime. 

\paragraph{Implications for PDR modeling}
\begin{table*}[h]
\begin{center}
\caption{ISM partitioning of measured line fluxes}
\label{tab:alpha_ism}
\begin{tabular}{l c c c c c c c c c}
\hline \hline
Line $i$& $L_i/L_{\rm FIR}$ & $L_i^{\rm M82}/L_{\rm FIR}^{\rm M82}$ & $\alpha_{i,\rm HII}$ & $\alpha_{i,\rm NLR}$ & \multicolumn{5}{c}{$\alpha_{i,\rm XDR}$} \\
\hline
            &                              &                                                                 &                                 &                                    & Case I\footnote{$n_{\rm H} = 10^5$ cm$^{-3}$, $R=600$ pc} & Case II\footnote{$n_{\rm H} = 5\times10^4$ cm$^{-3}$, $R=1000$ pc} & Case III\footnote{$n_{\rm H} = 10^5$ cm$^{-3}$, $R=500$ pc} & Case IV\footnote{$n_{\rm H} = 2\times10^5$ cm$^{-3}$, $R=350$ pc} & Case V\footnote{$n_{\rm H} = 2\times10^5$ cm $^{-3}$, $R = 1430$ pc}       
\\

$\rm [CII]158\mu$\rm m & $3.5\times10^{-3}$ & $3.2\times10^{-3}$ & 0.20                            & 0.20                             &  0.18 (0.13)\footnote{Parenthetical values have been calculated with the assumption of an M~82-like starburst.} &  0.24 (0.17) & 0.36 (0.26) & 0.20 (0.14) & 0.050 (0.035) \\ 
					&			&					   &			           &                                     &  \multicolumn{5}{c}{$\alpha_{i,\rm PDR}$} \\ 
					\cline{6-10}
					& 			& 					   & 			 	   &     				& Case I & Case II & Case III & Case IV & Case V \\
					&    			&					   &			           &					& 0.42 (0.47) & 0.27 (0.36) & 0.24 (0.34) & 0.40 (0.46) &  0.55 (0.57) \\
					&			&					   & 			   	  & 					& \multicolumn{5}{c}{$\alpha_{i,\rm XDR}$} \\
					 \cline{6-10}
					&			&					   & 			   	  & 					& Case I & Case II & Case III & Case IV & Case V \\
$\rm [OI]63\mu$\rm m & $3.9\times10^{-3}$ & $2.7\times10^{-3}$ & 0.0                            & 0.0                                   & 0.24 (0.17) & 0.11 (0.08) & 0.70 (0.51) & 0.77 (0.56) & 0.018 (0.013) \\
					& 			&					&				&					& \multicolumn{5}{c}{$\alpha_{i,\rm PDR}$} \\ 
					\cline{6-10}
					& 			&					&				&					& Case I & Case II & Case III & Case IV & Case V \\
					& 			&					& 				&					& 0.76 (0.83) & 0.89 (0.92) & 0.30 (0.49) & 0.23 (0.44) & 0.98 (0.99) \\  
$\rm [NII]122\mu$\rm m & $6.4\times10^{-4}$ & $5.7\times10^{-4}$ & 0.80 & 0.20 & - & - & - & - & -  \\
$\rm [OIII]52\mu$\rm m & $1.1\times10^{-3}$ & $9.7\times10^{-4}$ & - & - & - & - & - & - & -  \\
\hline
\end{tabular}
\end{center}
\end{table*}

Having determined the level of contribution to [CII]158$\mu$m and [OI]63$\mu$m fluxes from a range of XDRs capable of producing the observed CO emission, we can now examine the physical consequences of introducing these XDR contributions in the context of a composite ISM model where the gas is heated from both AGN activity and star formation. In particular, we would like to determine the effects of the X-ray component on physical parameters determined for the PDR gas by correcting the attributed PDR flux according to the following equations:
\begin{gather}
F_{\rm [OI],PDR} = (1 - \alpha_{\rm [OI],XDR})\times F_{\mathrm{[OI]}}  \\
 \begin{split}
F_{\rm [CII],PDR}= (1 &- \alpha_{\rm [CII],NLR} - \alpha_{\rm [CII],HII}\\
  &- \alpha_{\rm [CII], XDR}) \times F_{\mathrm{[CII]}}
\end{split}
\end{gather}

\begin{figure}
\includegraphics[width=0.45\textwidth]{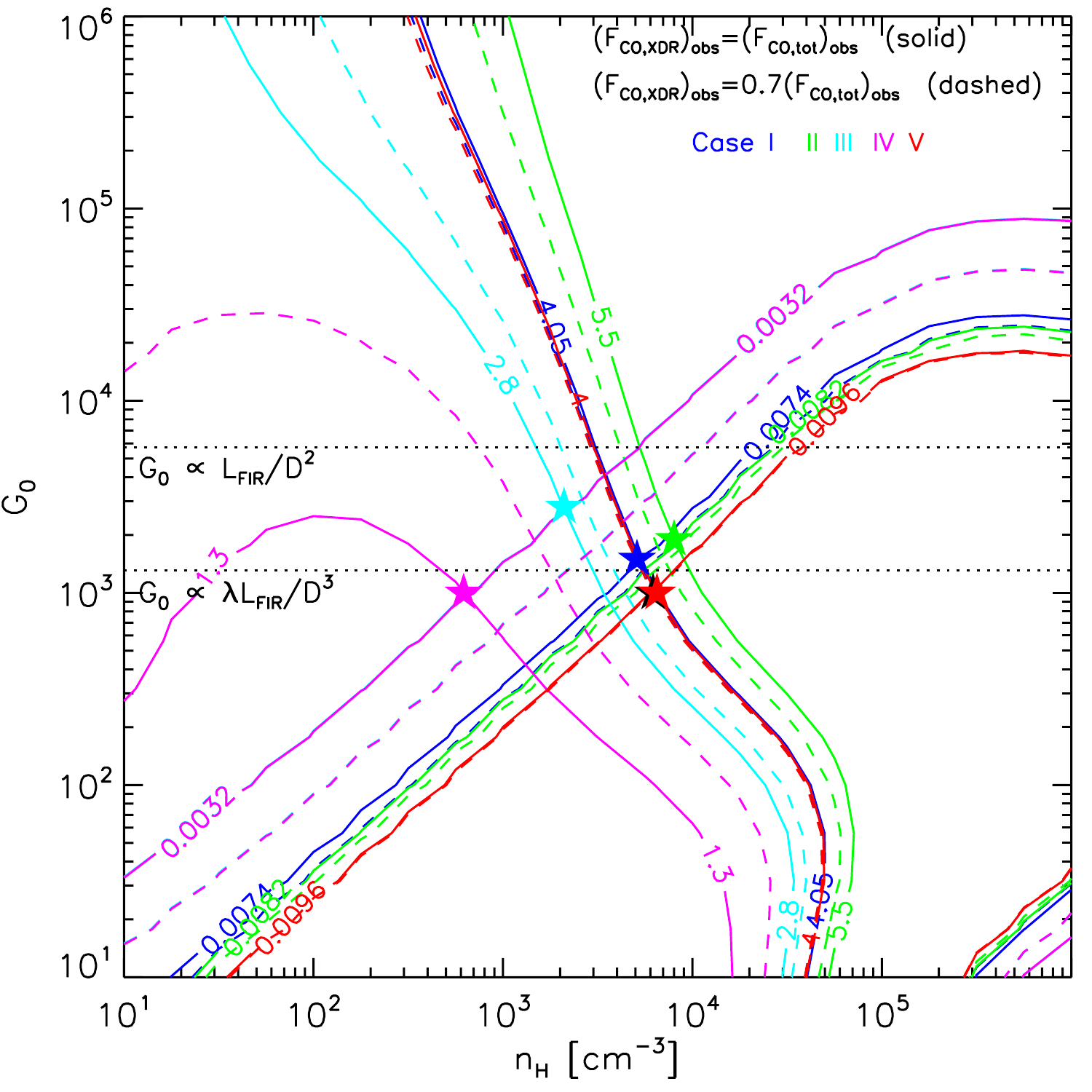}
\caption{PDR diagnostics for ISM models including a contribution from the CO-constrained XDR. Diagnostic ratios plotted are the same as in Figure~\ref{fig:K06diagnostic}: $F_{\rm[OI],PDR}/F_{\rm[CII],PDR}$ and $(F_{\rm[OI],PDR}+F_{\rm[CII],PDR})/F_{\rm FIR}$. (Recall that the curves for $(F_{\rm[OI],PDR}+F_{\rm[CII],PDR})/F_{\rm FIR}$ run from the lower-left to upper-right region of the plot, and the $F_{\rm[OI],PDR}/F_{\rm[CII],PDR}$ contours run from the upper-left to lower-right region.) Black star symbol at $G_0=10^3$ and $n_{\rm H}= 5.2\times10^3$ cm$^{-3}$ indicates the PDR solution for the NLR+HII+PDR model (i.e., excluding an XDR contribution). Colored star symbols indicate PDR solutions for the corresponding Cases of the same color defined in the legend. (Note that we artificially offset the positions of the red and black stars for better visibility; they indicate the same PDR solutions.) Constraints on $G_0$ from the analytic estimates are reproduced here as the dotted horizontal lines.}
\label{fig:xdr_pdr_diagnostics}
\end{figure}

Figure~\ref{fig:xdr_pdr_diagnostics} shows the resulting PDR diagnostic plots for each assumed starburst template, after contributions from XDR (Cases I--V), NLR, and star-forming HII region components have been subtracted from the measured line fluxes; the FIR continuum is unchanged because we do not expect significant contribution from ISM components other than PDRs. The ISM model with only NLR and HII region contributions (``NLR+HII+PDR"; indicated in Figure~\ref{fig:xdr_pdr_diagnostics} as the black star symbol at $G_0=10^3$ and $n_{\rm H}=6\times10^3$~cm$^{-3}$) serves as a reference point for assessing the effect of the CO-bright XDR on the derived PDR conditions. We find that, in general, the value of $G_0$ ($\sim10^3$) determined from the PDR models remains fairly robust to changes in the input $F_{\rm [CII],PDR}$ and $F_{\rm [OI],PDR}$ introduced by the CO-constrained XDR models, never falling outside the range of $G_0=$ 1--3$\times10^3$. In fact, the change in $G_0$ is maximized with respect to the NLR+HII+PDR model for XDR models with moderate densities and small radii (represented by Case III), where the [OI]63$\mu$m and [CII]158$\mu$m production in the corresponding XDRs can be significant: $\alpha_{\rm[OI],XDR}\sim0.7$ and $\alpha_{\rm[CII],XDR}\sim0.3$. This result brings the PDR solutions inferred from the FS line ratio diagnostics in better agreement with the analytic estimate for $G_0$ from Section~\ref{subsubsec:geo_gnot}. The PDR density is more sensitive to the given XDR model, with $n_{\rm H}$ spanning the range of $\sim0.7$--10$\times10^3$~cm$^{-3}$ for the Cases considered. The largest change to the PDR density---where $n_{\rm H}$ decreases to $\sim700$~cm$^{-3}$---on the other hand, occurs when adopting XDR models with small radii and high densities (Case IV), where the majority ($\sim55$--80\%, depending on the starburst template) of [OI]63$\mu$m flux is removed from the PDR and attributed to the XDR component.

\subsection{Alternative heating sources} \label{sec:mech_heating}

\paragraph{Shocks} Molecular gas excited by interstellar shock waves can also exhibit the high CO line-to-FIR continuum luminosity ratios characteristic of X-ray irradiated gas, due to the greater efficiency of gas heating relative to dust heating in the shock process. Among the HerCULES CO SLEDs plotted in Figure~\ref{fig:coSLEDs}, for example, NGC~6240---a known galaxy-galaxy merger with strong evidence for shock-excited molecular gas (cf. \citet{Meijerink2013} for CO and \citet{Lutz2003, Guillard2012} for H$_2$---is a clear outlier in terms of its observed CO-to-FIR continuum luminosity ratio---any given rotational transition above $J= 5$ carries on the order of $\sim10^{-4}$ of the system's FIR luminosity---and CO SLED shape. X-ray heating as a gas heating mechanism is disfavored in NGC~6240 because the CO emission does not originate close to either of the two AGN nuclei in the system, so that the incident X-ray flux on the 500-pc wide \citep{Tacconi1999} CO cloud is $\lesssim1$ erg s$^{-1}$ cm$^{-2}$ \citep{Komossa2003}. While there is no evidence, e.g. a disturbed morphology, to support an active merger in the Cloverleaf, there is no obvious reason to rule out shock excitation of CO in this case as interstellar shock waves can arise from phenomena related to the Cloverleaf's starburst and AGN, which are ample sources of mechanical e simulated extent ($\lesssim100$~pc) of the circumnuclear disk, accelerating ISM material and suppressing overall star-formation within this volume by factors of $\sim10$ \citep{Hopkins2016}. The total CO cooling is only 0.0049\% and 0.063\% of the total AGN bolometric power output and the starburst power output, respectively, while observations suggest that up to 5\% \citep{Cicone2014} of the total energy can go into mechnergy. Numerical simulations of quasar host galaxies show, for example, that mechanical feedback from AGN-driven winds dominates the entireanical processes for AGN. Of this 5\% put into mechanical energy, only $\gtrsim0.1$\% (or $\gtrsim1$\%) would be required to reproduce the total observed CO emission from the AGN (or star-formation).

\begin{figure}[b]
 \centering
 \includegraphics[width=0.47\textwidth]{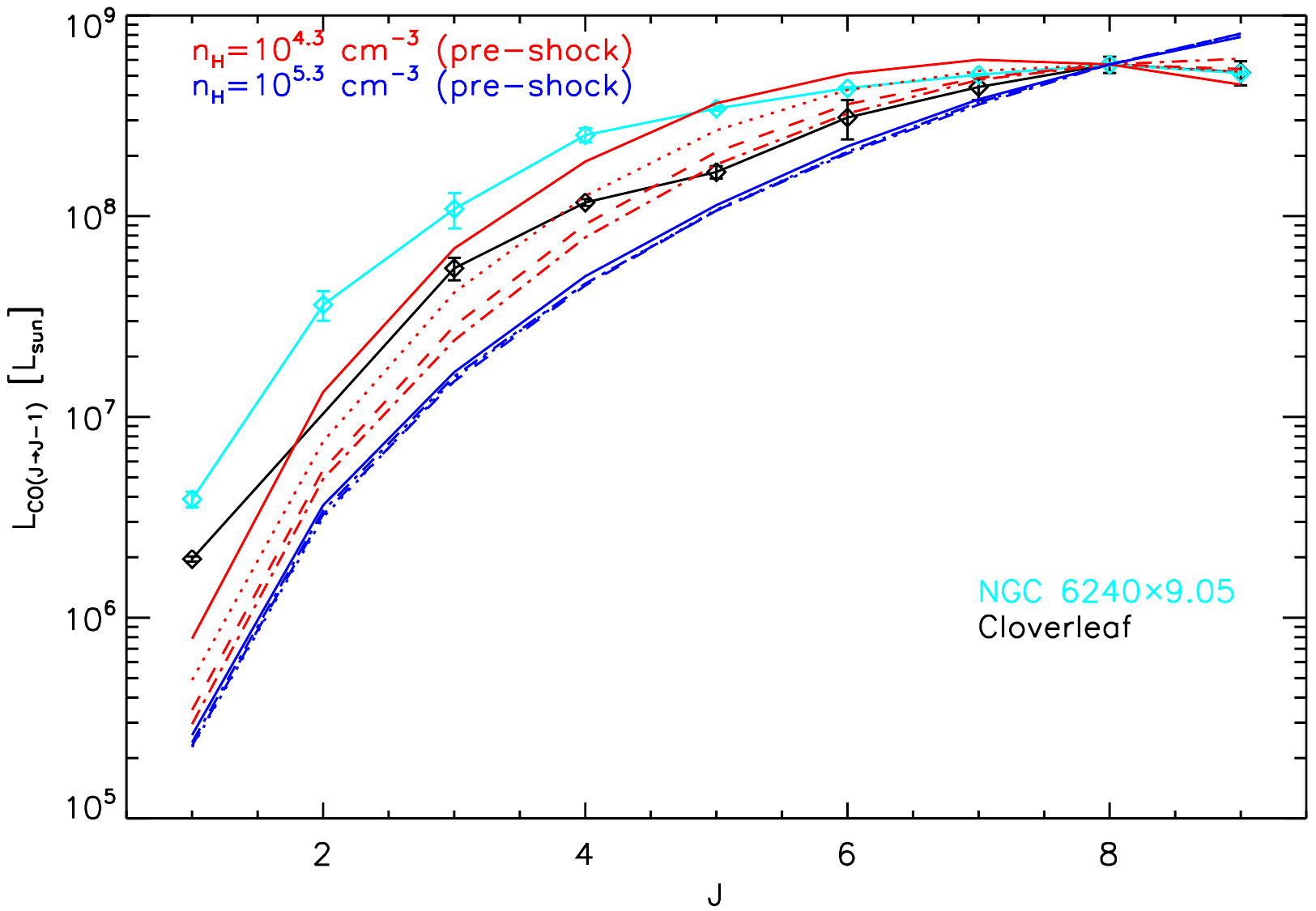}
 \caption{Predicted CO SLEDs for C-type shock models \citep{Flower2010} and observed CO SLED for the Cloverleaf (black curve). Red and blue curves correspond to pre-shock densities of $n_{\rm H} = 2\times10^4$ cm$^{-3}$ and $2\times10^5$ cm$^{-3}$, respectively. Different linestyles indicate shock velocities of 10 km $s^{-1}$ (\emph{solid}), 20 km $s^{-1}$ (\emph{dotted}), 30 km $s^{-1}$ (\emph{dashed}), and 40 km $s^{-1}$ (\emph{dot-dashed}). Models are normalized such that luminosities of the $J=8\rightarrow7$ transition match the observed luminosity for that transition in the Cloverleaf. For reference, the measured CO SLED for NGC~6240 (cyan curve)---which was reproduced in \citet{Meijerink2013} using a C-type shock model (not shown) with $n_{\rm H}=5\times10^4$~cm$^{-3}$ and $v_{shock} = 10$~km~s$^{-1}$---is also plotted. }
 \label{fig:shockmodels}
 \end{figure}

First we compare the CO excitation in the Cloverleaf to the predicted excitation from the shock models of \citet{Flower2010}. In their C-type shock models, which are considered here, transverse magnetic field strengths (in units of $\mu$G) for the pre-shock gas are given by $n_{\rm H}^{1/2}$, where density is in units of cm$^{-3}$. Figure~\ref{fig:shockmodels} shows the CO SLEDs for models of C-type shocks with pre-shock densities of $n_{\rm H} = 2\times10^4$ cm$^{-3}$ and $2\times10^5$ cm$^{-3}$, and velocities, $v_{shock}$, ranging from 10--40~km~s$^{-1}$. The CO SLEDs have been normalized to the luminosity of the $J=8\rightarrow7$ transition in the Cloverleaf. While it appears that there is no single shock model that reproduces the shape of the CO SLED for all measured transitions down to $J=1$, models with $n_{\rm H}=2\times10^4$~cm$^{-3}$ and velocities 30~km~s$^{-1}$ or 40~km~s$^{-1}$ provide a reasonable fit to the observed CO excitation for the transitions $J\ge5$. While B09 and \citet{Riechers2011clover} were able to reproduce the fluxes in all of the observed CO transitions down to $J=3$ and $J=1$, respectively, with single-component models, this fact alone does not rule out the scenario in which a range of physical conditions---such as multiple kinds of shocks with distinct velocities or multiple gas components with distinct densities---will contribute to the observed SLED. Therefore, we do not rule out shocks as a mechanism for CO excitation based solely on the inability of a model characterized by a single $n_{\rm H}$ and $v_{shock}$ to match the CO spectrum at $J < 5$. 

We also compare the global observed CO surface flux, 0.2~erg~s$^{-1}$~cm$^{-2}$, to the surface flux produced in the shock models. According to \citet{Flower2010}, the surface flux for the CO---where we have summed surface textbf{fluxes} reported for transitions $J=1$-17 to represent the total emission---produced in the model shocks with $n_{\rm H}=2\times10^4$~cm$^{-3}$ and shock speeds of 10, 20, 30, and 40~km~s$^{-2}$ are $4.0\times10^{-3}$, $9.1\times10^{-3}$, $1.3\times10^{-2}$, and $1.7\times10^{-2}$~erg~s$^{-1}$~cm$^{-2}$, respectively. Based on the incompatibility between the observed and predicted CO SLED shapes shown in Figure~\ref{fig:shockmodels}, we do not consider models with pre-shock densities of $2\times10^5$~cm$^{-3}$, but note that the surface flux is increased for these models, up to $9.2\times10^{-2}$~erg~s$^{-1}$~cm$^{-2}$ for $v_{shock}=40$~km~s$^{-1}$. Thus, the lower density shock models can only reproduce 2.0-8.5\% of the observed surface flux, or smaller percentages for $\phi_{\rm ff}<1$. If, however, $\phi_{\rm ff}>1$ in the Cloverleaf, then it may be possible to have a superposition of shocks in the same line-of-sight, thus increasing the resulting surface area of CO-emitting gas. For example, a total number of 8 million molecular gas clumps with average density $n_{\rm H}=2\times10^4$~cm$^{-3}$ and radius $r_{\rm clump}=0.8$~pc reproduces the observed molecular gas mass of $M_{\rm H_2}\sim8.5\times10^8$~M$_{\odot}$---where the value of $M_{\rm H_2}$ quoted here is the geometric mean of the ranges allowed by the likelihood analysis from B09---and yields a combined surface area of $6\times10^{44}$~cm$^{2}$. When distributed over a disk which has inner and outer radii of 180~pc (for the NLR radius inferred from F15) and 650~pc, this number and size of clumps gives an area filling factor of $\phi_{\rm ff} \sim 10$. Under the assumption that the shocks dissipate on scales equal to the clump radius, a series of $\sim10$ shock fronts with velocity $v_{shock}=40$~km~s$^{-1}$ propagating through a disk and sharing a single line-of-sight would produce a CO luminosity comparable to the observed value of $\sim1.3\times10^{43}$~erg~s$^{-1}$. We note that this value is likely the maximum CO luminosity possible in this case, since shocks could dissipate before reaching the edge of the molecular clump.

In addition to the shock models, we can consider the basic scaling relations of mechanical energy dissipation. In a turbulent medium where a large number of shocks supply sufficient mechanical energy to heat the molecular gas and drive its cooling, we can use the expression from \citet{Bradford2005}, namely, $L/M = 1.10 \times (v_{turb}/25 \ \mathrm{km \ s^{-1}})^3\times(1\ \mathrm{pc}/\Lambda)$~L$_{\odot}$~M$_{\odot}^{-1}$, to estimate the total molecular gas cooling provided by turbulent motions in the Cloverleaf ISM. In this expression, $L$ and $M$ are the total luminosity and mass in molecular gas, $v_{turb}$ is the velocity of turbulent motions, and $\Lambda$ is the physical dimension over which turbulence occurs, e.g, the molecular clump size. Lacking observations of the molecular line emission from H$_2$, the total luminosity of molecular line emission $L$ in the Cloverleaf is unknown, so we simply use the CO luminosity, $L_{\rm CO}$. This will underestimate the actual total molecular gas cooling per mass, but we note that for gas kinetic temperatures of 50--60~K, cooling via H$_2$ will be subdominant to CO. In that case, the total CO luminosity of $3.3\times10^{9}$~L$_{\odot}$ and molecular gas mass of $M_{\rm H_2}\sim8.5\times10^9$~M$_{\odot}$ yield a specific luminosity of $\sim0.39$~L$_{\odot}$~M$_{\odot}^{-1}$. We set $v_{turb} = v_{shock} = 30$ and $40$~km~s$^{-1}$ based on the SLED analysis above, and determine the size scale, $\Lambda$, that reproduces the observed cooling to be 5~pc and 10~pc, respectively, for $v_{turb}=30$~km~s$^{-1}$ and $40$~km~s$^{-1}$. We note that a total number of 33,000 (4,100) molecular gas clumps with an average density of $n_{\rm H} = 2.0\times10^4$~cm$^{-3}$ and typical radius of 5~pc (10~pc) in the Cloverleaf disk would amount to the measured molecular mass and have a column density, in each clump, of $N_{\rm H}=3.09\times10^{23}$~cm$^{-2}$ ($6.2\times10^{23}$~cm$^{-2}$), corresponding to $A_{\rm V}=150$ (300). With the same geometry considered above, this distribution of molecular clouds gives an area filling factor of $\phi_{\rm ff}=2.1$ ($1.1$). 
 
\paragraph{Cosmic-rays} Enhanced ionization rates of molecular hydrogen by cosmic-ray particles produced in supernovae may play a role in heating of molecular gas in starburst galaxies (e.g., \citet{Bradford2003} and \citet{Meijerink2011}), and we explore this possibility in the Cloverleaf. 

As outlined in \citet{Suchkov1993}, if one assumes that the rate of supernovae is proportional to a galaxy's SFR, it follows that the number density of cosmic-rays in a star-forming region, $n_{\rm CR}$, scales with the local SFR surface density. Further assuming that the cosmic-ray ionization rate, $\xi_{\rm CR}$, is proportional to $n_{\rm CR}$, then $\xi_{\rm CR}\propto n_{\rm CR}\propto\left(\frac{\rm SFR}{D^2}\right)$, and the ratio of cosmic-ray ionization rates for two starburst galaxies is equivalent to the ratio of their star formation rate surface densities. We then solve for the expected cosmic-ray ionization rate in the Cloverleaf, $\xi_{\rm CR}^{\rm clover}$, by scaling from the estimated ionization rate in M~82, $\xi_{\rm CR}^{\rm M82}$:
\begin{equation}
\xi_{\rm CR}^{\rm clover} = \xi_{\rm CR}^{\rm M82} \left(\frac{D_{\rm M82}}{D_{\rm clover}}\right)^2 \left(\frac{\rm SFR_{clover}}{\rm SFR_{\rm M82}}\right) \label{eq:cosmicray}
\end{equation}
The cosmic-ray density in M~82 has been measured as $\sim500$ times the density in the Galaxy \citep{Acciari2009}, so we scale $\xi_{\rm CR}^{\rm M82}$ by the same amount, such that $\xi_{\rm CR}^{\rm M82} = 500 \times$ (2--$7\times10^{-17}~\rm s^{-1})=$ 1--4$\times10^{-14}~\rm s^{-1}$. The lower and upper bounds in the allowable range indicate the Galactic cosmic-ray ionization rates as measured by \citet{Goldsmith1978} and \citet{vanDishoeck1986}, respectively. Cosmic-ray particles deposit roughly 20~eV of energy per H$_2$ ionization, so the amount of energy deposited into the molecular gas in M~82 is 3--10$\times10^{-25}$~erg~s$^{-1}$ per H$_2$ ionization. 

After comparing the star-formation rate surface densities for M~82 and the Cloverleaf, and applying Equation~\ref{eq:cosmicray}, we find that the cosmic-ray ionization rate is roughly 9 times greater in the Cloverleaf than in M~82. The energy deposition rate per H$_2$ ionization is then 3--9$\times10^{-24}$~erg~s$^{-1}$ for the Cloverleaf. The observed CO cooling implies that a supply of $8.5\times10^9$~M$_{\odot}$ ($\approx5\times10^{66}$~H$_2$~molecules) of molecular gas radiates $1.3\times10^{43}$~erg~s$^{-1}$, which equals $3\times10^{-24}$~erg~s$^{-1}$ per H$_2$ molecule,  so we conclude that cosmic-ray ionization could provide sufficient heating for the molecular gas in the Cloverleaf disk. This conclusion holds if we assume cosmic rays deposit a somewhat lower energy of $\sim13$~eV per H$_2$ ionization, as calculated in \citet{Glassgold2012}, for example.

We note that distinguishing between the enhanced cosmic ray- and XDR-heating scenarios will likely require observations of OH$^{+}$, H$_2$O$^{+}$, and H$_3$O$^+$ \citep{Meijerink2011, Rangwala2011}. We also point out that, in addition to starburst-related phenomena discussed above, cosmic rays can be produced in the magnetic fields in AGN \citep{Biermann2012, Laing2013, Meli2013}, though we do not presently examine the effects of AGN-supplied cosmic rays on the molecular gas in the Cloverleaf.

\section{Discussion} \label{sec:Discussion}
 
\subsection{Molecular clump sizes and spatial distribution} \label{subsec:clumps}

We have argued that the PDR gas and the CO-bright XDR gas are roughly coextensive in a $\sim1.3$~kpc-wide disk based on (1) the ability of predicted CO, [CII]158$\mu$m, [OI]63$\mu$m surface fluxes from theoretical XDR and PDR models to reproduce observed surface fluxes, and (2) similar distributions of the spatially resolved CO(7-6) and 122$\mu$m continuum maps (with assumption that the 122$\mu$m continuum is produced by SF). This scenario implies that the SF, which provides UV-heating for the PDRs, is not inhibited by exposure to strong X-ray fluxes. It is important to note, however, that the \emph{Herschel} line fluxes represent aggregate emission across the Cloverleaf system, and thus do not preclude the possibility that the CO-bright XDR gas is actually at smaller radius---which would imply that it is illuminated by stronger X-ray fields than the PDRs. While the construction of a more detailed model of the disk structure is not justified given the limitation of the current dataset, there are plausible scenarios that would allow for SF to occur in gas exposed to moderate to high X-ray fluxes, which we now examine.

For instance, it is possible that SF is occurring in regions which are more highly shielded than the CO-bright XDR, as may occur, for example, in an azimuthally non-uniform X-ray field. Alternatively, the star-forming regions could be shielded by local gas, such as in a dense envelope of molecular gas. The hydrogen gas density required to attenuate X-ray flux by a factor of 22 or more is readily achieved by the gas densities indicated by CO observations, which are in excess of  $10^4$ cm$^{-3}$, and possibly as high as 10$^5$ cm$^{-3}$. If we let $n_{\rm H} = 3\times10^4$ cm$^{-3}$, then a spherical molecular gas clump that could potentially host SF would have a column of attenuating hydrogen of $N_{\rm H,att}=6.5\times10^{23}$~cm$^{-2}$ at a clump radius of 7~pc. The presence of $\approx7,900$ such clumps would total the observed molecular gas mass of the Cloverleaf, namely, $M_{\rm H_2} \sim 8.5\times10^9$ M$_{\odot}$, and gives rise to $\phi_{\rm ff}=1$ when distributed in the disk of outer radius 650~pc and inner radius 180~pc. While the area filling factor may be high for this clump distribution, the volume filling factor would be only $\sim6\%$, providing a relatively unobstructed path for X-rays to reach each molecular clump once it leaves the foreground screen in front of the AGN. Smaller (larger) clumps have sub- (greater than) unity area filling factors, which lead to insufficient (excessive) X-ray illuminated gas to produce the observed CO, at least in the existing model. 
 
\subsection{Comparison with high-redshift and local systems}

Co-spatial SF and SMBH accretion at sub-kpc scales has been suggested for another molecular gas-rich quasar at higher redshift, APM~08279+5255 ($z=3.9$). In that system, the CO gas is also spatially resolved to be concentrated within a short distance of the AGN---$R = 550$ pc---and similarly emits a large CO surface flux that is best-matched to an XDR component of comparable size to the full extent of the molecular disk \citep{Bradford2011}. Observations of high-order rotational transitions of water molecules in APM~08279, coupled with high dust temperatures throughout the disk that are uncharacteristic of XDRs, suggest that SF is ongoing in the XDR as well, if radiative pumping from FIR photons of re-processed starlight from dust is responsible for exciting the water as suggested by \citet{Bradford2011} and \citet{vanderWerf2011}. 

At lower redshift, the local ULIRG Markarian~231 most closely resembles the Cloverleaf in terms of its FIR luminosity and the dominance of the AGN in its overall energetics. Mrk~231 also exhibits an XDR contribution to the CO emission, but, unlike the Cloverleaf, this XDR is more centrally concentrated to the inner $\sim160$ pc of the 550 pc molecular disk; the majority of the low-$J$ CO emission is emitted in spatially extended PDR components at $R>160$ pc \citep{vanderWerf2010}. The LIRG and composite starburst/AGN NGC~1068 is another local example where XDRs have been identified as the excitation mechanism for CO-emitting molecular gas \citep{HaileyDunsheath2012, Spinoglio2012ngc1068}, although the starburst component in that galaxy is modeled as being distributed in a ring at larger radius than the CO-bright XDR, which is distributed in a circumnuclear disk.

Additionally, we note that the measured [CII]-to-FIR continuum luminosity ratio of $3\times10^{-3}$ in the Cloverleaf is consistent with observations of this ratio in a sample of quasars and submm galaxies with $L_{\rm FIR}>10^{12}$~L$_{\odot}$ at $z\sim$ 1--3 as compiled, e.g., in \citet{CarilliWalter2013} (cf. their Figure~6). The observed ratios at this redshift range are consistent with moderate FUV fluxes of order $G_0\sim 10^3$, characteristic of kpc-scale star formation.

\section{Conclusions}
Observations of the dominant PDR cooling lines [CII]158$\mu$m and [OI]63$\mu$m have allowed us to assess the physical conditions---parametrized by the gas density and the incident FUV flux---prevalent in atomic gas heated by stellar populations in the Cloverleaf. After subtracting expected contributions to the measured [CII]158$\mu$m flux from the NLR and HII regions associated with star-formation, we find that K06 PDR models with $n_{\rm H}=5.6\times10^3$ cm$^{-3}$ and $G_{0}=10^3$ reproduce the observed diagnostic flux ratios $F_{\rm [OI]}/F_{\rm [CII]}$ and $\left(F_{\rm [CII]} + F_{\rm [OI]}\right)/F_{\rm FIR}$. These moderate conditions, however, contradict the findings from PDR models derived exclusively from CO diagnostic ratios, which instead favor gas densities $n_{\rm H} > 10^5$ cm$^{-3}$ and imply greater emission in [OI]63$\mu$m and less emission in [CII]158$\mu$m than what is seen in the Cloverleaf.  Thus, we conclude that UV heating from local star formation is not sufficient to explain both the observed atomic line and CO luminosities, and suggest that additional X-ray heating components from the AGN are required and may well dominate the molecular gas heating. 

Based on predicted [CII]158$\mu$m and [OI]63$\mu$m XDR surface fluxes for a range of XDR models that reproduce the measured CO surface flux, we have identified a set of viable XDR models which do not overproduce the observed [OI]63$\mu$m or [CII]158$\mu$m fluxes, and which can be combined with PDR models to inform global ISM conditions in the Cloverleaf molecular disk. Due to broad likelihood distributions of the derived molecular gas densities, temperatures, and thermal pressures from previously published results of CO modeling, as well as uncertainties in the area filling factor of the gas and the precise contribution of the Cloverleaf starburst to the total CO emission, we have considered a correspondingly expansive parameter space of combined XDR and PDR models in our analysis.

The general picture of the Cloverleaf ISM that emerges from our composite model has the [CII]158$\mu$m and [OI]63$\mu$m line emission produced primarily within PDRs and HII regions associated with star-formation in the host. The actual star-formation sites are embedded within a $\sim1.3$~kpc diameter clumpy molecular gas disk for which X-rays dominate the heating (with possible contributions from turbulence or cosmic rays). If densities and X-ray fluxes are large, then the X-ray heating may be responsible for the majority (up to 70-80\%) of observed [OI]63$\mu$m, but will not contribute substantially to the [CII]158$\mu$m emission. While the X-rays may or may not be the dominant heating source, energetics and geometry of the source make it difficult to consider a scenario where the X-ray heating is not important---we have a situation in which star-formation is ongoing in molecular material which is immersed in a strong X-ray radiation field. Similarly, the cosmic ray energy density due to the star-formation itself is also many times that of even starburst galaxy nuclei. Such large uniform bulk heating of the gas may change the properties of the star-formation relative to the standard paradigm. For example, B09 suggest that bulk heating may result in an effective increased Jeans mass, and so suppress the formation of low-mass stars. We suggest that this is a topic for careful theoretical study.
\\\\
The authors would like to thank Nanyao Lu for help assessing the noise level in the \emph{Herschel}-FTS spectra of the Cloverleaf. We thank Carl Ferkinhoff for discussions related to the [NII]122$\mu$m data from ALMA. We thank Lee Armus and Tanio D\'{i}az-Santos for helpful discussions and comments on the manuscript, and for providing us with results from an analysis of [CII]158$\mu$m and PAH emission in GOALS galaxies. We would also like to thank Aaron Evans, Brent Groves, J.D. Smith, and Fabian Walter for helpful discussions related to this work. BDU acknowledges support from the NASA Graduate Student Research Program fellowship and NSF AST 1455151.

\appendix

\begin{figure}
\centering
\begin{tabular}{c}
\includegraphics[width=0.45\textwidth]{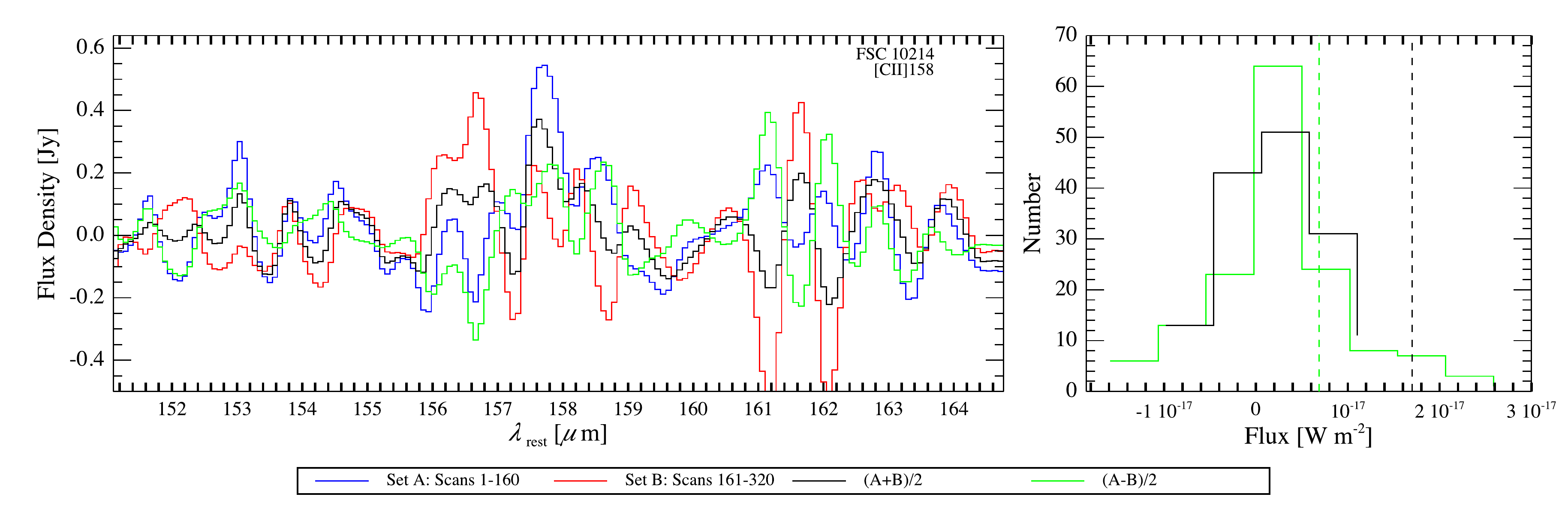} \\
\includegraphics[width=0.45\textwidth]{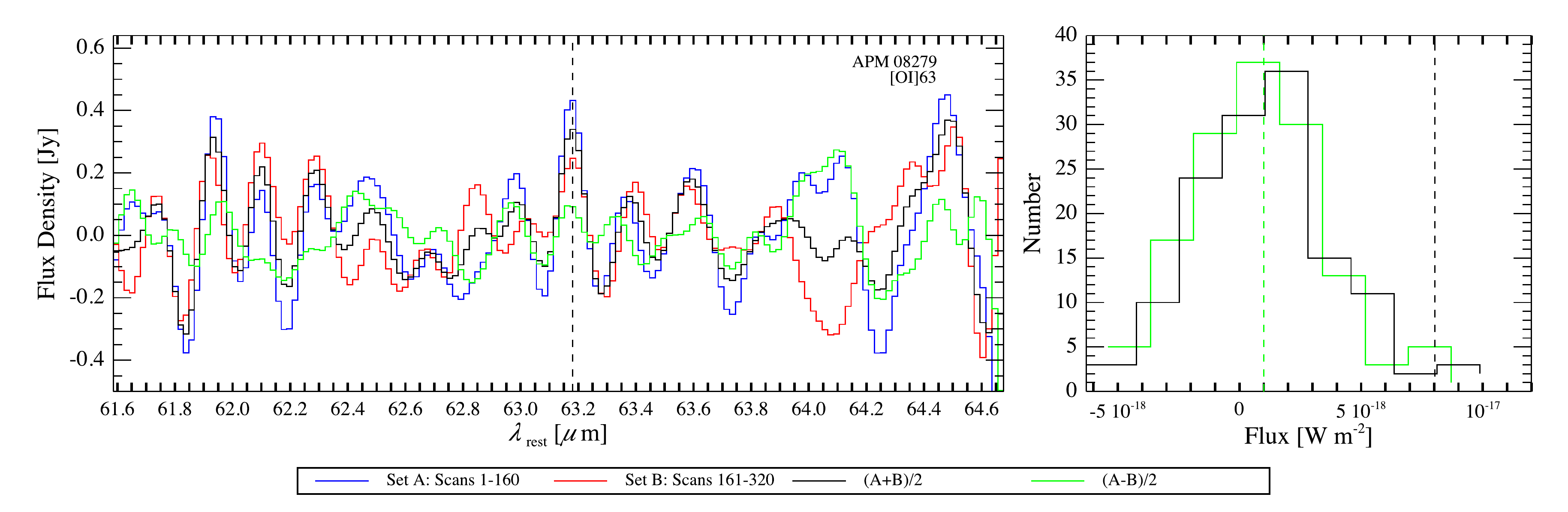} \\
\end{tabular}
\caption{Same as Figure~\ref{fig:jackknives_hist}, with SPIRE-FTS spectra for [CII]158$\mu$m in FSC~10214+4724 (top panel) and [OI]63$\mu$m in APM~08279+5255 (bottom panel).}
\label{fig:fsc_apm_ciioi63}
\end{figure}

In addition to the Cloverleaf, the high-redshift quasars APM~08279+5255 ($z=3.91$), FSC~10214+4724 ($z=2.29$), and MG~0751 ($z = 3.20$) were also targeted for observations with \emph{Herschel} SPIRE-FTS in the \texttt{OT1\_mbradfor\_1} program . We present in this appendix the results of a search for [CII]158$\mu$m and [OI]63$\mu$m emission in these sources. 

For FSC~10214+4724, we report a $3\sigma$ upper limit on the apparent [CII]158$\mu$m flux as $<2.2\times10^{-17}$~W~m$^{-2}$. We have have performed similar jackknife tests and flux uncertainty estimates as in Section~\ref{sec:observations} on the portion of the SPIRE-FTS spectrum containing the expected [CII]158$\mu$m emission, and show the original and jack-knifed spectrum, as well the histogram of fluxes for each wavelength position in the spectrum, in Figure~\ref{fig:fsc_apm_ciioi63} (upper panel). Given the apparent IR luminosity of this source, $6\times10^{13}$~L$_{\odot}$ \citep{Evans2006}, this upper limit implies $F_{\rm [CII]}/F_{\rm FIR}  < 3.9\times10^{-3}$, which is consistent with the observed value in the Cloverleaf.

For APM~08279+5255, we report a $3\sigma$ upper limit on the apparent [OI]63$\mu$m flux as $<1.6\times10^{-17}$~W~m$^{-2}$. Original and jack-knifed spectrum, and flux histogram, are shown in Figure~\ref{fig:fsc_apm_ciioi63} (lower panel). The apparent FIR luminosity of APM~08279+5255 is $2\times10^{14}$~L$_{\odot}$ \citep{Weiss2007}, implying $F_{\rm [OI]}/F_{\rm FIR}<3.0\times10^{-3}$, which is consistent with the ratio found for the Cloverleaf. The redshifted [CII]158$\mu$m line for this source falls outside of the spectral coverage of SPIRE-FTS.

No detections or upper limits are reported for MG~0751.

\end{document}